\documentclass[a4paper,fleqn,usenatbib]{mnras}
\usepackage{newtxtext,newtxmath}
\usepackage[T1]{fontenc}
\usepackage{ae,aecompl}

\usepackage{graphicx}	% Including figure files
\usepackage{amsmath}	% Advanced maths commands
\usepackage{amssymb}	% Extra maths symbols
\usepackage{lscape}	% Extra maths symbols
\usepackage{multirow}
\usepackage{xcolor}
\usepackage{ulem}
\title[$^{12}$CO, $^{13}$CO and C$^{18}$O in dusty star-forming galaxies]
{VALES VI: ISM enrichment in star-forming galaxies up to z$\sim$0.2 
using $^{12}$CO(1-0), $^{13}$CO(1-0) and C$^{18}$O(1-0) line luminosity ratios}

\author[M\'endez-Hern\'andez et al.]
{H. M\'endez-Hern\'andez, $^{1}$\thanks{E-mail: hugo.mendez@postgrado.uv.cl}
E. Ibar,$^{1}$
K. K. Knudsen, $^{2}$
P. Cassata, $^{3, 4}$
M. Aravena, $^{5}$
\newauthor
M. J.  Micha{\l}owski, $^{6}$
Zhi-Yu Zhang, $^{7}$
M. A. Lara-L\'opez, $^{8}$
R. J. Ivison, $^{19}$
\newauthor
P. van der Werf, $^{10}$
V. Villanueva, $^{11}$
R. Herrera-Camus, $^{12}$
T. M. Hughes $^{13, 1, 14, 15}$
\\
$^{1}$Insituto de F\'isica y Astronom\'ia,
Universidad de Valpara\'iso, Avda. Gran  Breta\~na 1111, 2340000
Valpara\'iso, Chile\\ 
$^{2}$Department of Space, Earth and
Environment, Chalmers University of Technology, Onsala Space
Observatory, SE-439 92 Onsala Sweden.\\ 
$^{3}$Dipartimento di Fisica e
Astronomia Galileo Galilei, Universit\`a degli Studi di Padova Vicolo
del$L'$Osservatorio 3, 35122 Padova Italy\\ 
$^{4}$ INAF Osservatorio
Astronomico di Padova, vicolo dell'Osservatorio 5, I-35122 Padova,
Italy\\ 
$^{5}$N\'ucleo de Astronom\'ia, Facultad de Ingenier\'ia y
Ciencias, Universidad Diego Portales, Av. Ej\'ercito 441, Santiago,
Chile \\ 
$^{6}$ Astronomical Observatory Institute, Faculty of
Physics, Adam Mickiewicz University, ul.~S{\l}oneczna 36, 60-286
Pozna{\'n}, Poland \\ 
$^{7}$  School of Astronomy and Space Science,
Nanjing University, Nanjing 210093, China Key Laboratory of Modern
Astronomy \\ and Astrophysics (Nanjing University), Ministry of
Education, Nanjing 210093, China \\ 
$^{8}$ DARK, Niels Bohr Institute,
University of Copenhagen, Lyngbyvej 2, Copenhagen DK-2100, Denmark\\
$^{9}$ European Southern Observatory, Alonso de C\'ordova, 3107,
Vitacura, Santiago 763-0355, Chile \\ 
$^{10}$ Leiden Observatory,
Leiden University, P.O. Box 9513, NL-2300 RA Leiden, The Netherlands\\
$^{11}$ Department of Astronomy, University of Maryland, College Park,
MD 20742, USA\\ 
$^{12}$ Astronomy Department, Universidad de
Concepci\'on, Barrio Universitario, Concepci\'on, Chile \\ 
$^{13}$ Chinese Academy of Sciences South America Center for Astronomy,
China-Chile Joint Center for Astronomy, \\ Camino El Observatorio
\#1515, Las Condes, Santiago, Chile\\ 
$^{14}$ CAS Key Laboratory for
Research in Galaxies and Cosmology, Department of Astronomy,
University of Science and Technology of China, \\ Hefei 230026, China\\ 
$^{15}$ School of Astronomy and Space Science, University of
Science and Technology of China, Hefei 230026, China\\}

\date{Accepted 2020 June 29. Received 2020 June 01; in original form 2019 December 04}

\pubyear{2020}
\graphicspath{{Figures2/}}
\begin{document}
\label{firstpage}
\pagerange{\pageref{firstpage}--\pageref{lastpage}}
\maketitle
\begin{abstract}

We present Atacama Large  Millimeter/sub-millimeter Array (ALMA)
observations  towards 27 low-redshift ($0.02< z<0.2$) star-forming
galaxies taken from the Valpara\'iso  ALMA/APEX Line Emission Survey
(VALES). We perform stacking analyses of the $^{12}$CO($1-0$),
$^{13}$CO($1-0$) and C$^{18}$O($1-0$) emission lines to explore the
$L'$\,($^{12}$CO($1-0$)\,)/$L'$\,($^{13}$CO($1-0$))) (hereafter 
$L'$\,($^{12}$CO\,)/$L'$\,($^{13}$CO))  and
$L'$\,($^{13}$CO($1-0$)\,)/$L'$\,(C$^{18}$O($1-0$)) (hereafter
$L'$\,($^{13}$CO\,)/$L'$\,(C$^{18}$O) line luminosity ratio dependence
as a function of different global galaxy parameters related to the
star formation activity.  The sample has far-IR luminosities
$10^{10.1-11.9}\,$L$_{\sun}$ and stellar  masses of
$10^{9.8-10.9}$M$_{\sun}$ corresponding to  typical star-forming and
starburst galaxies at these redshifts. On average we find a
$L'$\,($^{12}$CO\,)/$L'$\,($^{13}$CO) line luminosity  ratio value of
16.1$\pm$2.5. Galaxies with evidences of possible merging  activity
tend to show higher $L'$\,($^{12}$CO\,)/$L'$\,($^{13}$CO) ratios  by a
factor of two, while variations of this order are also found in 
galaxy samples with higher star formation rates or star formation
efficiencies. We also find an average 
$L'$\,($^{13}$CO\,)/$L'$\,(C$^{18}$O) line  luminosity ratio of
2.5$\pm$0.6, which is in good agreement with those  previously
reported for starburst galaxies. We find that galaxy samples with high
$L_{\text{IR}}$, SFR and SFE  show low
$L'$\,($^{13}$CO\,)/$L'$\,(C$^{18}$O) line luminosity ratios  with
high $L'$\,($^{12}$CO\,)/$L'$\,($^{13}$CO) line  luminosity ratios,
suggesting that these trends are produced by selective enrichment of
massive stars in young starbursts.

\end{abstract}

\begin{keywords}
methods: statistical.
techniques: interferometric.
galaxies: star formation.
galaxies: ISM.
\end{keywords}

%%%%%%%%%%%%%%%%%%%%%%%%%%%%%%%%%%%%%%%%%%%%%%%%%%

%%%%%%%%%%%%%%%%% BODY OF PAPER %%%%%%%%%%%%%%%%%%

\section{Introduction} 
Stars are mostly formed within Giant Molecular Clouds (GMCs), cold
dense regions of the interstellar medium (ISM), which are
characterized by  high densities ($n_{\text{H}_{2}}$ > 10$^{4}$
cm$^{-3}$; \citealt{Gao04,Bergin07}) and low temperatures (10--20\,K;
\citealt{Evans99}) that favour the formation of stars. In these
regions, the most abundant molecule is molecular hydrogen, H$_{2}$,
however its lack of a permanent electric dipole makes it difficult to
observe  in emission. After H$_{2}$, the next most abundant molecule
is carbon monoxide, $^{12}$C$^{16}$O (hereafter CO), which easily
emits photons from low level rotational transitions in similar ISM
conditions as those in which the H$_{2}$ molecule resides. Therefore,
the CO emission from low-$J$ rotational transitions have become the
workhorse tracer of the H$_{2}$ gas mass in the local Universe and
beyond \citep{Bolatto13}.

 Since the CO emission is mostly optically thick within GMCs,   
 optically thin CO isotopologues are usually used to look deeper into 
 the densest regions of GMCs.  Since $^{12}$C, $^{16}$O and their 
 isotopes, $^{13}$C and $^{18}$O,  are mainly products of primary and
 secondary stellar nucleosynthesis  processes, they are powerful
 tracers of the evolutionary state of a galaxy, and represent
 excellent tools to characterize the physical conditions and the 
 chemical processes of the ISM \citep{Wilson94,Milam05,Romano17}. 
 \cite{Narayanan14} showed that along side gas density and
 temperature, the  optical depth from $low$-J CO lines is well
 correlated with the star formation  rate surface density of GMCs.
 Moreover, it is possible to trace different  stellar nucleosynthesis
 scenarios, by comparing $^{12}$CO, $^{13}$CO and  C$^{18}$O abundance
 variations. For example, \cite{Henkel93} showed that dense regions
 that recently  experienced a star formation event are expected to
 have higher abundances  of C$^{18}$O and $^{12}$CO compared to
 $^{13}$CO. The  $^{12}$C/$^{13}$C abundance ratio reflects the
 relative degree of primary to secondary nucleosynthesis processing,
 while the $^{18}$O/$^{16}$O abundance ratio traces differences in the
 Initial Mass Function (IMF) \citep{Milam05,Romano17}. In practice
 albeit the optical depth effects, we could assume that the
 $^{12}$C/$^{13}$C and $^{18}$O/$^{16}$O abundance ratios can be
 traced  by the molecular $I$\,($^{12}$CO\,)/$I$\,($^{13}$CO) and
 $I$\,($^{13}$CO\,)/$I$\,(C$^{18}$O) line intensity ratios
 respectively

After the first detection of $^{12}$CO and its isotopologues in the
Milky Way  \citep{Wilson70,Penzias71}, several works have repeated
their detection in nearby galaxies
(\citealt{Rickard75,Encrenaz79,Rickard85,Young86}). More recently,
several works have proven successfully the usage of 
$I$\,($^{12}$CO\,)/$I$\,($^{13}$CO) and
$I$\,($^{13}$CO\,)/$I$\,(C$^{18}$O) line ratio in nearby galaxies
\citep{Sliwa17,JimenezD17,Cormier18,Sliwa17,Brown19}, and lensed 
high-redshift galaxies \citep{Henkel10,Danielson13,Spilker14,Zhang18}.

An environmental dependence for $^{12}$C/$^{13}$C has been shown by
\citet{Alatalo15}, who found that 17 Early Type Galaxies (ETG) located
in the Virgo cluster and groups, showed a line intensity ratio about
two times lower than field galaxies. They proposed three different
scenarios in which the observed variations could be explained: an
extra low-mass stellar enrichment taking place in Virgo cluster
galaxies, an increased mid-plane pressure effects of the intracluster
medium (ICM) or the survival of only the densest clumps of molecular
clouds as galaxies  enter the ICM. Additionally, \citet{Davis14}
showed a systematic dependence of the
$I$\,($^{12}$CO\,)/$I$\,($^{13}$CO) line intensity ratio on the star
formation rate surface density ($\Sigma_{\text{SFR}}$) and the
molecular gas surface density ($\Sigma_{\text{H}_{2}}$) using a sample
of nearby starburst and early-type galaxies. They suggest that the
observed correlations  are caused by the combined action of massive
stars heating and/or  inducing turbulence in the gas phase on those
galaxies with higher $\Sigma_{\text{SFR}}$. 

Recent works have reported  $I$\,($^{13}$CO\,)/$I$\,(C$^{18}$O) line
intensity ratios for different galaxy types.  \cite{Danielson13}
showed a low $I$\,($^{13}$CO\,)/$I$\,(C$^{18}$O)  line intensity ratio
($\sim$1) in a high redshift lensed galaxy   suggesting the presence
of a significant fraction of  high-mass stars. \cite{Sliwa17} reported
a simultaneous high  $I$\,($^{12}$CO\,)/$I$\,($^{13}$CO) ($\gg$60)
intensity ratio with a low $I$\,($^{13}$CO\,)/$I$\,(C$^{18}$O)
($\lesssim1$)  intensity ratio consistent with an ISM enrichment by
the presence of a young starburst, a top-heavy IMF or their combined
action. \cite{JimenezD17} presented a
$I$\,($^{13}$CO\,)/$I$\,(C$^{18}$O) line intensity ratio dependency
with  $\Sigma_{\text{SFR}}$ and galactocentric distance in nine nearby
spiral galaxies due to the selective enrichment of the ISM by massive
stars.  More recently, \cite{Zhang18} showed high
$I$\,($^{12}$CO\,)/$I$\,($^{13}$CO) line intensity ratios with a
simultaneous low $I$\,($^{13}$CO\,)/$I$\,(C$^{18}$O) line intensity
ratios in four gravitationally lensed sub-millimetre galaxies (SMGs)
at $z\sim$2-3, and claimed this to be caused by a change of the IMF
where there is a higher number of massive stars in high-$z$ starburst
galaxies than in typical galaxies.

For galaxies beyond the Local Universe, the observation of faint
emission lines as $^{13}$CO or C$^{18}$O is usually challenging.  The
abundances of $^{13}$CO and C$^{18}$O are typically  50 and 500 times
lower than $^{12}$CO \citep{JimenezD17} and  their flux density ratios
usually range between 20 to 100  for I($^{12}$CO)/I($^{13}$CO)  and
between 20-140 for I($^{12}$CO)/I(C$^{18}$O) 
\citep{Aalto91,Casoli92b,Konig16,Sliwa17}.  For individual detections
in nearby local ULIRGs, $^{13}$CO and C$^{18}$O  observations need to
be at least four times deeper than $^{12}$CO observations  to yield
line detections \citep{Sliwa17,Brown19}. In this work, we propose an
alternative way to overcome sensitivity limitations by stacking the
signals of the $^{12}$CO($1-0$), $^{13}$CO($1-0$) and C$^{18}$O($1-0$)
lines ($\nu_{^{12}\text{CO(1-0)}}=115.271$\,GHz,
$\nu_{^{13}\text{CO(1-0)}}=110.201$\,GHz, and 
$\nu_{\text{C}^{18}\text{O(1-0)}}=109.782$\,GHz rest-frame
frequencies, respectively) from individual star-forming galaxies to
produce a statistically robust study for the content of these 
isotopologues up to $z=0.2$.\\ 

This paper is structured as follows. In Section \ref{sec:Data}, we
present the Atacama Large  Millimetre/sub-millimetre Array (ALMA) data
used in this work. Section~\ref{sec:analysis} details the way we
modelled the stacked $^{12}$CO, $^{13}$CO  and C$^{18}$O fluxes as
well as their errors. Section~\ref{sec:Res} presents the
$^{12}$CO($1-0$)/$^{13}$CO($1-0$) and
$^{13}$CO($1-0$)/C$^{18}$O($1-0$) (hereafter 
$L'$\,($^{12}$CO\,)/$L'$\,($^{13}$CO) and 
$L'$\,($^{13}$CO\,)/$L'$\,(C$^{18}$O)) line luminosity ratio
measurements  and their dependence as a function of global galaxy
parameters, while Section~\ref{sec:Disc} presents the discussion.
Finally, our conclusions are shown in Section~\ref{Sec:Concl}.
Throughout this work, we assume a $\Lambda$CDM cosmology adopting the
values H$_{0}$\,=\,70\,km\,s$^{-1}$\,Mpc$^{-1}$, $\Omega_{\textrm{M}}$
= 0.3 and $\Omega_{\Lambda}$= 0.7 for the calculation of luminosity
distances and physical scales.

\section{Data}\label{sec:Data}
\subsection{Sample}\label{sec:Sample}

In this work, we present $^{13}$CO(1-0) and C$^{18}$O(1-0) line
measurements for 27 and 24 galaxies, respectively, which were
previously detected by {\it Herschel} in [C\,{\sc ii}] 
(\citealt{Ibar15}) and with ALMA in $^{12}$CO
(\citealt{Villanueva17}).  The sample is part of the Valpara\'iso
ALMA/APEX Line Emission Survey (VALES; \citealt{Villanueva17,Cheng18})
 designed to characterize the CO emission line of low-$J$ transitions
from typical star-forming and starburst galaxies up to z\,$=$\,0.35.
The parent population comes from dusty galaxies taken from the
equatorial fields of the {\it Herschel} Astrophysical Terahertz Large
Area Survey ($H$-ATLAS; \citealt{Eales10}.) Galaxies were selected
using a spectroscopic redshift at  0.02$<z<0.2$, and a {\it Herschel}
detection  near the peak of the spectral energy distribution (SED) of
a  normal star-forming galaxy  ($S_{160\mu\text{m}}>$150$\mu$Jy).  All
galaxies have an unambiguous optical counterpart in the  Sloan Digital
Sky Survey  (SDSS; \citealt{Adelman08}), have high-quality spectra
from the Galaxy and Mass Assembly survey (GAMA
\footnote{http://www.gama-survey.org/}; \citealt{Liske15, Driver16}
z$_{\text{QUAL}}$ $\geq$ 3),  and show a Petrosian SDSS radii smaller
than 15$\arcsec$ (see \citealt{Ibar15} for more details).

\subsection{ALMA $^{13}$CO($1-0$) and C$^{18}$O($1-0$) observations}\label{ssec:13COObs}

\begin{table*}
\centering
\caption{\label{tab:table1}
New ALMA $^{13}$CO($1-0$) and C$^{18}$O($1-0$) observations 
(Project ID: 2013.1.00530.S) presented in 
this work. 'PWV' is the average precipitable water vapour estimate for the
observations. All data were taken using 32 12-m ALMA antennas. 
One observation taken on 24 January 2015 failed to run through the pipeline 
due to unknown reasons, so we have arbitrarily removed it from this work. 
Note that $^{12}$CO($1-0$) observations can be found in 
\protect\cite{Villanueva17}.
}
\begin{tabular}{lccccc}
\hline
Target names & 
Observation & 
Flux & 
Bandpass & 
Phase & 
PWV \\
 HATLAS & 
 Date & 
 Calibrator & 
 Calibrator & 
 Calibrator & 
 [mm] \\
\hline\hline
J085340.7+013348,
J085405.9+011130     & \multirow{2}{*}{2015 January 24 (1/3)} & \multirow{8}{*}{Ganymede}  & \multirow{8}{*}{J1058+0133} &
\multirow{8}{*}{J0909+0121}& \multirow{2}{*}{5.9} \\ %v2.103 v2.137
J085356.4+001255,
J083601.5+002617     & & & & & \\ %v2.170 v2.23

J085112.9+010342, 
J090949.6+014847  &\multirow{2}{*}{2015 January 24 (2/3)}& & & & 
\multirow{2}{*}{5.2}\\ %v2.235 v2.26
J085450.2+021208,
J091205.8+002655  & & & & & \\ %v2.38 v2.42

J085346.4+001252,
J084428.4+020350 &\multirow{2}{*}{\sout{2015 January 24 (3/3)}}& & & &\multirow{2}{*}{4.5}\\ %v2.45 v2.52
J090005.0+000446,
J090532.6+020222 & & & & & \\%v2.55 v2.58

J085111.4+013006, 
J083745.1-005141, &\multirow{2}{*}{2015 January 25}& & & & 
\multirow{2}{*}{4.5}\\ %v2.60 v2.66
J085828.6+003813,
J085233.9+013422  & & & & & \\%v2.80 v2.87

\hline

J084350.8+005534, 
J083831.8+000044      & \multirow{2}{*}{2015 January 23 (1/2)} & \multirow{4}{*}{J0750+125} & \multirow{4}{*}{J0909+0121} &
\multirow{4}{*}{J0901-0037} & \multirow{2}{*}{3.8} \\ %v2.102 v2.111
J084305.1+010855
J084907.1-005138      & & & & & \\ %v2.167 v2.175

J084217.9+021223
J084139.6+015346 &\multirow{2}{*}{2015 January 23 (2/2)}& & & & \multirow{2}{*}{3.9}\\ %v2.232 v2.299
J085748.0+004641,
J084428.4+020657 & & & & & \\ %v2.48 v2.77
\hline

J090750.0+010141, 
J085836.0+013149      & \multirow{2}{*}{2015 January 23}&    \multirow{2}{*}{J0854+201}  & \multirow{2}{*}{J0750+1231} &
\multirow{2}{*}{J0901-0121} & \multirow{2}{*}{3.8} \\ %v2.107 v2.76
J084630.9+005055      &    & & & & \\ %v2.90
\hline
\end{tabular}
\end{table*}

Observations with ALMA in band-3 were performed as part of project
2013.1.00530.S (P.I.\, E.\ Ibar), targeting the redshifted
$^{12}$CO($1-0$), $^{13}$CO($1-0$) and C$^{18}$O($1-0$) emission lines
for 27 VALES galaxies. The $^{12}$CO($1-0$) observations reached a
root mean square (rms) of  2\,mJy\,beam$^{-1}$ at  a spectral
resolution of 30\,km\,s$^{-1}$ and are presented in
\cite{Villanueva17}. The simultaneous $^{13}$CO($1-0$) and
C$^{18}$O($1-0$) observations were taken between 23 and 25 January
2015, in compact configuration  (maximum  baseline of $\sim$300\,m) 
with precipitable water  vapour (PWV) conditions in the range
$\sim$4--6\,mm. The observational strategy consisted of grouping
sources in terms of redshift, such that we could observe all 27
galaxies using just three spectral setups (each one using four
spectral windows to cover 7.5\,GHz of bandwidth). The grouped sources
are shown in  Table~\ref{tab:table1}, including the different
executions performed by ALMA to reach the requested sensitivity.
Unfortunately,  the spectral setup missed C$^{18}$O($1-0$) coverage in
three galaxies.

Data reduction and imaging were performed using the same procedure as
in \cite{Villanueva17}, where we developed a common pipeline within
the Common Astronomy Software Applications (CASA version 4.4.0) to
process all of the science goals. Each source was imaged with the {\sc
tclean} task using a natu- ral weighting. This yielded a restoring
beam between 3$"$ and 4$"$, nevertheless for the purposes of this
work, we fixed the restoring beam to a common value, at $4\farcs5$,
for all sources.   The $^{13}$CO($1-0$) and C$^{18}$O($1-0$)
observations reached  rms\ noise of  0.9\,mJy\,beam$^{-1}$\ at
30\,km\,s$^{-1}$  channel width  ($\sim2\times$ deeper than $^{12}$CO
observations).  We note that $\sim$110 GHz continuum emission is
undetected at $5\sigma$  significance in all sources down to a rms\
noise of 4$\mu$Jy\,beam$^{-1}$.

\section{Analysis} 
\label{sec:analysis}

Out of the 27 galaxies, 26  have been previously spectrally detected
at $>5\sigma$ significance (signal-to-noise ratio: SNR) in
$^{12}$CO($1-0$) \citep{Villanueva17}. The $^{13}$CO line was visually
inspected  for any individual detection. There were no confident
$^{13}$CO($1-0$) emission lines from individual spectra for any of the
27 galaxies. Nevertheless, using the information of the $^{12}$CO line
widths as priors, we created moment-0 maps by collapsing the cube
around $\pm1\times$ FWHM$_{^{12}\text{CO}}$ of the expected $^{13}$CO
frequencies. In the collapsed images we identify 7 galaxies with SNR >
5.  The remaining 21 galaxies have not been detected above a 5$\sigma$
significance in their moment  zero maps. Table \ref{tbl:13COIndvDet}
shows the SNRs, velocity integrated line flux densities and
luminosities of these individual $^{13}$CO detections. With respect to
the C$^{18}$O emission line, we do not identify any  detection in the
spectra nor in the individual moment-0 maps using the same approach
mentioned above.

\begin{table} 
\caption{$^{13}$CO($1-0$)
detections from collapsed spectral images using $\pm$FWHM $^{12}$CO km\,s$^{-1}$ 
line width around $^{13}$CO($1-0$) expected frequencies. 
(col 1) ID taken from \protect\cite{Villanueva17}, 
(col 2) observed signal-to-noise ratio in moment 0maps, 
(col 3) velocity integrated line flux densities} with error
measurements, 
(col 4) $^{13}$CO($1-0$) luminosity with error measurements.
\label{tbl:13COIndvDet}
\begin{tabular}{|l|c|c|c}
\hline
 \multicolumn{1}{|c|}{ID} &
 \multicolumn{1}{c|}{SNR$_{^{13}\text{CO}}$}  &
 \multicolumn{1}{c|}{S$_{^{13}\text{CO}} \Delta$v}  &
 \multicolumn{1}{c|}{$L'_{^{13}\text{CO}}$} 
 \\
 \multicolumn{1}{c|}{HATLAS}  &
 \multicolumn{1}{c|}{}  &
 \multicolumn{1}{c|}{mJy\,km\,s$^{-1}$} &
 \multicolumn{1}{c|}{K km\,s$^{-1}$pc$^{2}$}
 \\

\hline
J090949.6+014847 & 5.6 & 490 $\pm$  90  & 68.0 $\pm$ 12.1 \\%&BC  0.018 \\v2.26-49 
J085346.4+001252 & 5.7 & 225 $\pm$  40  &  2.9 $\pm$  0.5 \\%&D   0.050 \\v2.45-51 
J084139.6+015346 & 6.1 & 206 $\pm$  33  &  5.4 $\pm$  0.9 \\%&-   0.073 \\v2.299-55 
J084350.8+005534 & 6.2 & 458 $\pm$  73  & 11.9 $\pm$  2.0 \\%&DBC 0.072 \\v2.102-56 
J083831.8+000044 & 6.4 & 147 $\pm$  22  &  4.3 $\pm$  0.7 \\%&DBC 0.078 \\v2.111-59 
J085748.0+004641 & 5.9 & 343 $\pm$  58  &  8.9 $\pm$  1.5 \\%&M   0.071 \\v2.48-64 
J090633.6+001526 & 7.0 & 710 $\pm$  100 &  9.6 $\pm$  1.4 \\%&DB  0.051 \\v2.58-50 
 \hline\end{tabular}
\end{table}

Different techniques have been proposed to detect the emission of
faint emission lines, falling below the detection limits.  For example
\citet{Loomis18} proposed a matched filtering method that uses a
previously identified  high signal-to-noise emission line as a kernel
for filtering the  $uv$ signal and thereby facilitate the detection of
any contiguous faint emission line. Similarly, \citet{Yen16}  proposed
an image-plane line detection technique tailored  to boost the SNR of
faint emission lines in keplerian disks. An independent approach has
been the development of stacking techniques. This has been successful
to detect the combined signal of faint emission  coming from multiple
objects of the same population over the electromagnetic  spectrum,
including the  X-ray \citep{Bartelmann03,Rodighiero15,Yang18}, UV
\citep{Berry12,Rigby18}, infrared \citep{Dole06,Duivenvoorden20}, 
submm \citep{Webb03,Knudsen05,Ibar13,Millard20}, and radio
\citep{Miller13,Bera18,Perger19} regimes. Moreover, stacking
techniques have proven to be a robust method for line and continuum
detections of high redshift galaxies
\citep{Scoville07,Lehmer07,Schinnerer07,Miller08}. In order to compute
 stacked line ratios, in this study we explore  three different
techniques: two of them in the image plane: i) stacking all the
moment-0 maps, ii) stacking all the frequency channels of all sources
following a channel by channel basis,  and additionally by iii)
stacking the individual $uv$-plane average signals.

\subsection{Image stacking}\label{sec:ImageStacking}

\subsubsection{2D-moment-0 stacking}\label{sec:M0Stacking} Based on
the previously detected $^{12}$CO line  widths and intensity peaks, we
 collapsed the datasets to create moment-0 maps. For this,  we measure
$^{12}$CO line widths  using 20 km\,s$^{-1}$ channel for all galaxies.
Using the {\sc immoments} \textrm{CASA} task, we collapse each galaxy
cube to create moment-0 maps for all $^{12}$CO, $^{13}$CO and
C$^{18}$O datasets,  around ($\pm1\times$ FWHM$_{^{12}\text{CO}}$) the
$^{12}$CO, $^{13}$CO and C$^{18}$O expected frequencies. We visually
inspected all of the 27 collapsed  $^{12}$CO images to correct for any
possible spatial offsets with respect to the intensity peak. Such
offsets exist; optical and submm observations trace the stellar and
the  molecular gas content of galaxies respectively, and thus the
location of  the peaks do not necessarily match. Given that the
reference coordinates of our ALMA observations were obtained from
optical images, we apply astrometric corrections to our $^{12}$CO
intensity maps, in order to correct any discrepancy between optical
and $^{12}$CO images. These corrections are on average of the order of
$\sim1$\farcs$4$  in random directions (smaller than the synthesized
beam of 4$\farcs$5). Finally, using a stacking code that we developed,
these images were stacked to obtain final collapsed signals reaching
rms values of  108 \,mJy\,beam$^{-1}$ \,km\,s$^{-1}$ for the $^{12}$CO
line  and  18 \,mJy\,beam$^{-1}$ \,km\,s$^{-1}$ for the $^{13}$CO and
C$^{18}$O emission lines.  We note that these stacked values are
$\sim$\,5 times deeper than individual moment-0 images.

To extract velocity integrated line flux densities from the stacked
signals,  we create $30"\times30"$ stamps and model the  sources with
a 2D Gaussian profile, assuming that the stacked signals are 
point-like  with a FWHM of 4$\farcs$5.

\subsubsection{3D-image stacking}\label{sec:3DStacking}

\begin{figure} \centering
\includegraphics[width=\columnwidth]{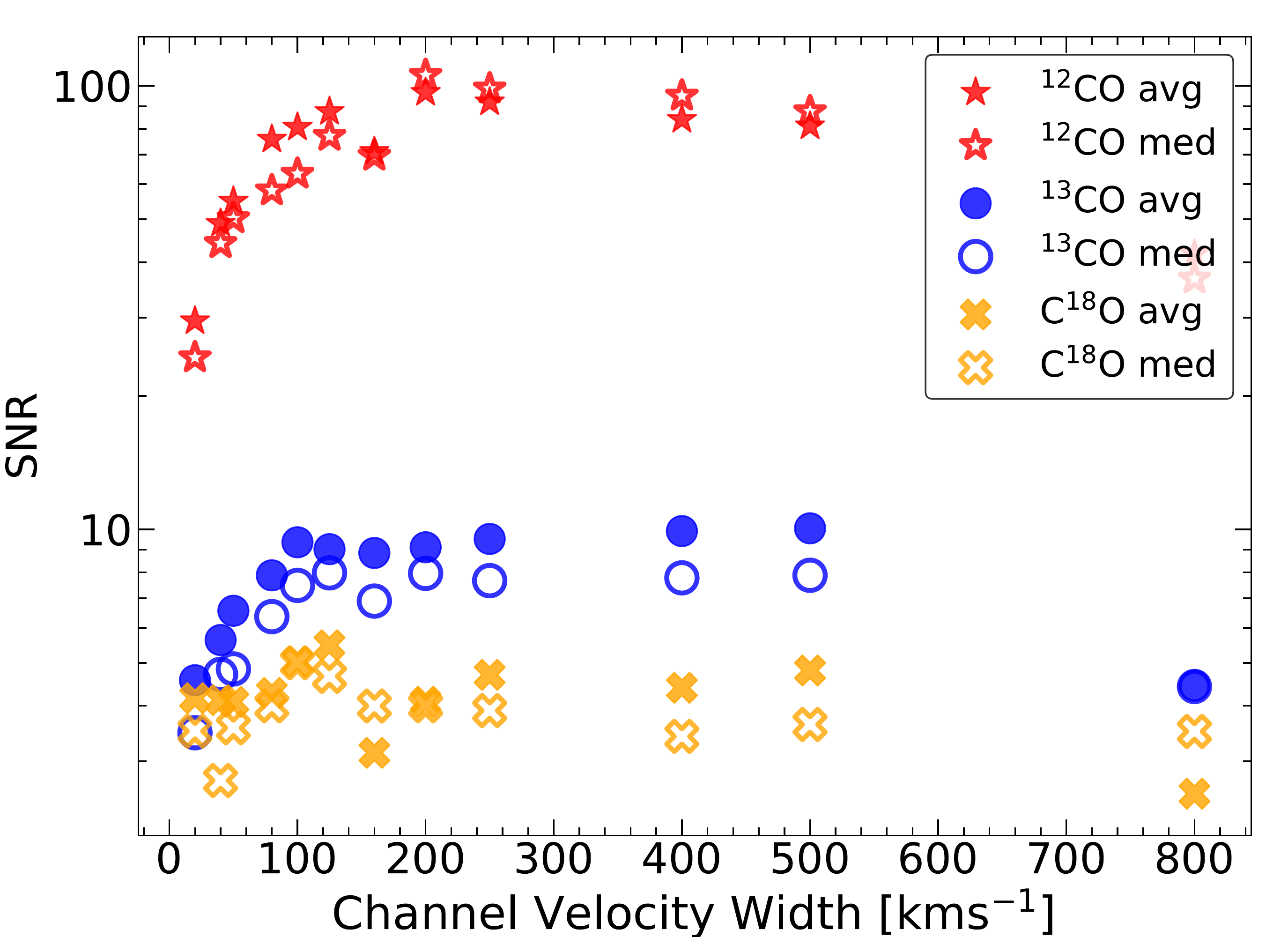}
\caption{Stacked signal-to-noise ratios (SNRs) obtained from 
$^{12}$CO (stars), $^{13}$CO (circles) and C$^{18}$O (crosses) using
different velocity channel widths, from 20\,km\,s$^{-1}$ to 
800\,km\,s$^{-1}$. Filled symbols correspond to average SNR values
while empty symbols correspond to median SNR  values. These
measurements are used to identify the best spectral  channel width for
3D stacking (see \S~\ref{sec:3DStacking}).}\label{fig:12CO13COPNR}
\end{figure}

In this approach, for processing the datacubes, we consider a  common
spectral channel width for $^{12}$CO, $^{13}$CO and C$^{18}$O 
emission lines. To determine the best common spectral channel width to
use, we kept in mind the idea of optimizing the SNR of the final
stacked data cubes detections of both lines. For this, datacubes for
all of the 27 galaxies were created using \textrm{CASA} task {\sc
tclean} assuming different velocity bin widths: [20, 40, 50, 80, 100,
125, 160, 200, 250, 400, 500, 800] km\,s$^{-1}$.  After this, we
obtained a 3D stacked cube by combining  the individual galaxy cubes
following a channel by channel and pixel by pixel basis. We then
obtained the SNR by measuring the peak at the central image pixel and
central velocity channel (0km\,s$^{-1}$). We measured the noise in the
image excluding the central region. Peak flux densities were recorded
for both the mean and median stacked cubes with different velocity
width bins and for $^{12}$CO, $^{13}$CO and C$^{18}$O data sets. 
Figure \ref{fig:12CO13COPNR} shows  this central channel SNR as a
function of channel velocity width. We find that the $^{12}$CO signal
maximizes at channel width of 200\,km\,s$^{-1}$ for both the median
(SNR$\sim$105.7) and mean  (SNR$\sim$96.6). The stacked $^{13}$CO line
maximizes at 125km\,s$^{-1}$ (SNR$\sim$8.0) and 500km\,s$^{-1}$
(SNR$\sim$10.1) for the median and mean respectively. Finally, the
C$^{18}$O line maximizes at 125km\,s$^{-1}$ for both the median
(SNR$\sim$4.7) and the mean (SNR$\sim$5.5) stacks.  Based on these
results we decided to use a common  spectral channel width  of
125\,km\,s$^{-1}$ for $^{12}$CO, $^{13}$CO and C$^{18}$O to image all 
data cubes in order to optimize the signal-to-noise ratio in the final
stacked images.  We note that the same astrometric offsets used for
moment-0 stacking, and described in Section \ref{sec:M0Stacking} have
been applied here, whilst velocity offsets based on the peak observed
in $^{12}$CO  are applied to all of the independent cubes  in order to
re-centre the signal. These offsets originate from small  differences
between the optical and submm redshifts which trace different phases
of the ISM.

\begin{figure} \centering
\includegraphics[width=\columnwidth]{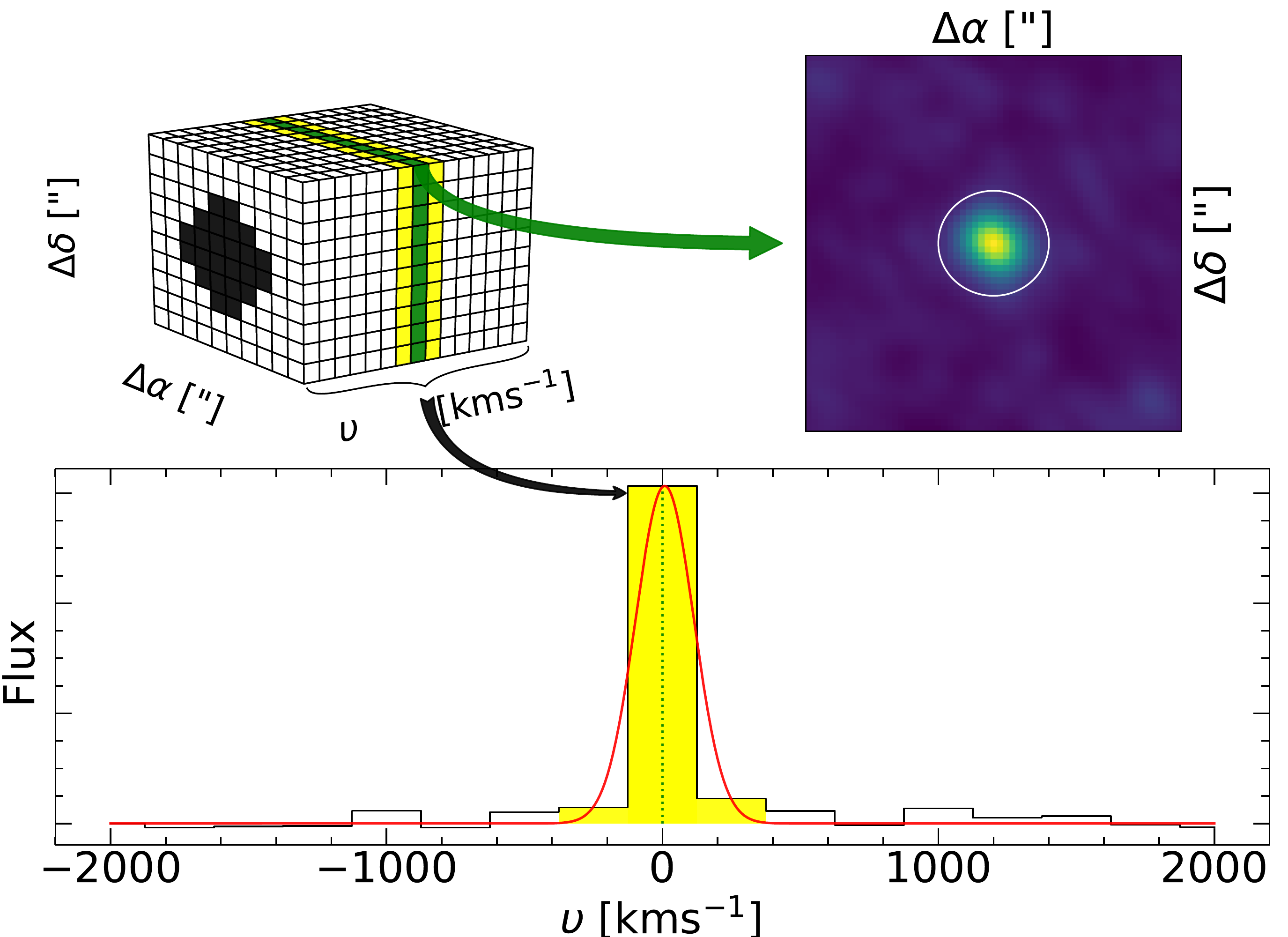} \caption{Velocity
integrated flux density measurements on a 3D stacked data cube. Upper
left panel: 3D stacked image cube showing the central channel (green)
at which the peak of the line  is located, the channels covered by the
fit FWHM$_{f}$ line width highlighted in yellow and the 15$\arcsec$
radius aperture (black)  used to generate the spectral line profile
shown below.  The bottom panel shows the spectral line profile (solid
black line) and shows  the 1D gaussian fit (red line) to obtain the
stacked velocity line  width FWHM$_{f}$ highlighted in yellow. The
upper right panel shows the  central channel where a 2D Gaussian
profile is fit (white) to obtain  the amplitude of source. Both line
width and amplitude are  used to compute the velocity integrated line
flux densities.} \label{fig:3DStack-Scheme1} \end{figure}

Using a common channel width of 125\,km\,s$^{-1}$, a  spectral
coverage of $\pm$2000km\,s$^{-1}$, and a restoring beam with a FWHM of
4$\farcs$5, we created the individual datacubes which are then stacked
to  get a cube containing the average $^{12}$CO, $^{13}$CO and
C$^{18}$O signals. In order to measure velocity integrated line flux
densities, we first created spectral line profiles using a fixed
aperture of 15$\arcsec$ radius ($\gtrsim$3$\times$synthesized beam), 
centred at the source position. Thereby, we fitted a 1D Gaussian
profile  to obtain the global stacked velocity  line width  FWHM$_{f}$
(see Fig. \ref{fig:3DStack-Scheme1} lower panel).  Hence, we took the
central channel (0km\,s$^{-1}$, where the peak in the spectral line
profile is located) to fit a 2D  Gaussian profile assuming that the
signal is point-like (see Fig. \ref{fig:3DStack-Scheme1} upper right
panel). Finally, the  amplitude of the 2D Gaussian  fit together with
the line width,  is used to calculate velocity integrated line flux
densities.

\subsubsection{Systematic Errors}\label{sec:3DStacking-SysErrors} 

In order to compute the systematic errors for our  stacked velocity
integrated line flux density measurements, we ran Monte-Carlo
simulations using data cubes with the same physical scales  (pixel
size, synthesised beam, primary beam) as those covered by the ALMA
Band-3 observations.

We model each source as point-like (spatially) using a  2D circular
Gaussian profile and spectrally by a 1D Gaussian profile centered  at
0\,km\,s$^{-1}$. We simulate a spectral coverage of
$\pm$2000\,km\,s$^{-1}$.  These sources are added to a random,
normally-distributed background noise  that has been convolved to the
scale of the synthesised beam. 

\begin{figure}
\includegraphics[width=\columnwidth]{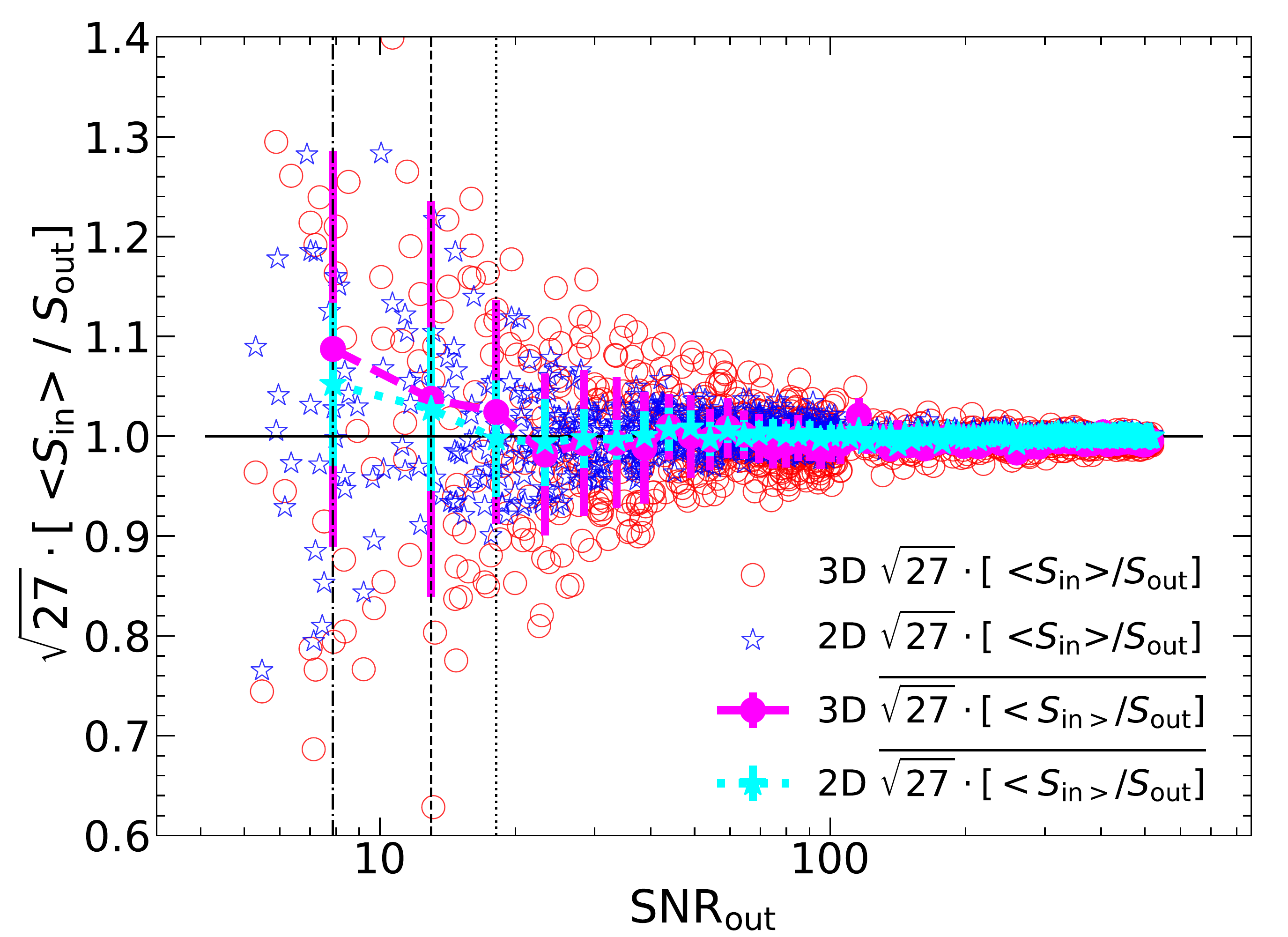}
\caption{ Simulated accuracy of the velocity integrated flux density
measurements (<$S_{\text{in}}$>/$S_{\text{out}}$) after stacking 27
galaxies.  Stacked detections with high SNR$_{\text{out}}$ have
clearly  better accuracy for the velocity integrated flux density
measurements. <$S_{\text{in}}$> refers to the average velocity
integrated flux densities of the simulated sources used for the
stacks, while $S_{\text{out}}$ refers  to the measured velocity
integrated flux density on the composite stacked images. 3D stacks are
shown as open circles, 2D moment-0 stacks are  shown as open stars and
filled symbols indicate the average  
$\overline{<S_{\text{in}}>/S_{\text{out}}}$ binned by 
SNR$_{\rm{out}}$. Vertical lines indicate the location of our
$^{12}$CO, $^{13}$CO and C$^{18}$O SNR$_{\text{out}}$ stacked
detections at $\sim$6 (dashed-dotted), $\sim$13  (dashed) and $\sim$18
(dotted) where 3D-image stacks show systematic errors of 8\%, 3\% and
2\% while 2D-moment-0 stacks show 5\%, 3\% and 1\%  respectively.}
\label{fig:SNR-FLUX-SYTNH} \end{figure}

To simulate the stacking, we take 27 data cubes with sources at fixed
signal-to-noise ratios (<SNR$_{\text{in}}$>) and fixed velocity  line
widths. We stacked them and compute the velocity integrated line flux
densities as  described in Sections \ref{sec:M0Stacking}  and
\ref{sec:3DStacking}.  We repeat this process 1000 times,  where
amplitudes and velocity widths are simulated to take fixed values
between 0 and 100 times the rms\ and line FWHMs  between 50 and
500\,km\,s$^{-1}$, respectively. The extracted velocity integrated 
line flux densities ($S_{\text{out}}$) are measured and compared to
the input values.  Figure~\ref{fig:SNR-FLUX-SYTNH} shows the
<$S_{\text{in}}$>/$S_{\text{out}}$  ratio of 1000 simulated stacked 
data cubes for both, 2D-moment-0 (open stars) and 3D-image  (open
circles) stacking methods. The average binned 
$\overline{<S_{\text{in}}>/S_{\text{out}}}$ ratios (filled symbols in
Fig.~\ref{fig:SNR-FLUX-SYTNH})  as a function of SNR$_{\text{out}}$ of
the composite stack images in the range between zero and 520
($=100\times\sqrt{27}$).  We note that the stacked images have
SNR$_{\text{out}}$  that are $\sim\sqrt{27}$ times larger than the
average <SNR$_{\text{in}}$> of the individual images given by  Poisson
factor gained by the stacking approach.  For example, if the $^{13}$CO
stack has a measured SNR$_{\text{out}}\sim$9, then this is produced by
individual datacubes with an average  <SNR$_{\text{in}}$>$\sim$1.73.
Figure \ref{fig:SNR-FLUX-SYTNH} shows vertical lines indicating three
different SNRs values  $\sim$6, $\sim$13 and $\sim$18 at which our
3D-stacks show  systematic errors of 8\%, 3\% and 2\% while
2D-moment-0 stacks show 5\%, 3\% and 1\%. These results show how
typically 2D-stacks show smaller systematic errors than 3D-stacks.
These differences tend to become negligible at SNRs>15.

\subsection{$uv$ stacking}\label{sec:UVStacking}

Interferometric telescopes provide data that samples the brightness
distribution of an observed source in Fourier space, where a point
measurement per integration time is provided by a pair of antennas.
The location of every point (visibility) in the Fourier space
($uv$-plane) is determined by the separation of a pair of antennas as
they trace the track of the source in the Fourier space during
integration. The imaging process considers a deconvolution which
assumes interpolations made on the $uv$-plane that could lead to
artifacts in the extracted images due to the intrinsic non-continuous
sampling nature of interferometric datasets \citep{Condon16}.
Interferometric stacking analyses are usually performed using these
reconstructed images. \cite{Lindroos15} developed \textit{stacker}
\footnote{https://www.oso.nordic-alma.se/software-tools.php}, a tool
which directly stacks interferometric continuum datasets in the
$uv$-plane  providing typical signal-to-noise ratios which are 20\%
higher compared with continuum image stacking. \textit{Stacker} was
designed to perform the stacking analysis for continuum $uv$-data,
therefore we mimic continuum maps as the average single channel maps
of the $^{12}$CO, $^{13}$CO  and C$^{18}$O emission lines intensity
data of the galaxies. To create  individual single channel maps we use
the \textrm{CASA} task {\sc split}  to obtain an average $uv$-data of
the channels  around the $^{12}$CO line observed frequency
($\pm1\times$  FWHM$_{^{12}\text{CO}}$). As described in
Section~\ref{sec:M0Stacking}  $^{12}$CO line widths were measured
using a 20km\,s$^{-1}$ resolution,  and these widths were also used to
create the individual single channel  maps for the $^{13}$CO and
C$^{18}$O datasets. Similarly to previous  approaches, we note that we
have applied the same astrometric offset  corrections to generate the
single channel $uv$ maps. To measure  velocity integrated line flux 
densities from the $uv$ stacks,  we create images using CASA task {\sc
tclean} following a similar approach  as the images used for 2D and
3D-image stacking procedures. Then as in  2D-stacks, we model  the
sources with a 2D Gaussian profile and measured  the velocity
integrated line flux densities from $30"\times30"$  stamps. 

\subsection{The differences between the stacking approaches}
\label{sec:StackingComp}

In this section we discuss the stacking approaches  described above in
order to decide the most suitable one for our work. As mentioned
before, each method is based on different assumptions, therefore a
direct comparison is not entirely trivial. For example, the images
obtained from 2D-stacking are generated using CASA  task {\sc
immoments} which basically sums the intensities of the channels around
$\pm1\times$\,FWHM$_{^{12}\text{CO}}$ for the $^{12}$CO observed
frequency, while the $^{13}$CO and C$^{18}$O lines are blindly
extracted at the expected frequencies using the  derived $^{12}$CO
redshifts. The 3D approach concentrates mainly on highlighting the
intensities from the central channel  of data cubes, where the
velocity peak of the flux density profile is located. The case for the
stacks obtained from a $uv$ approach are  constructed starting from 
CASA task {\sc split} which averages the $uv$ intensities of the
channels where the lines are located, these channels are exactly the
same as those channels used to create the moment-0 maps  for
2D-stacking.

\begin{figure}
\includegraphics[width=\columnwidth]{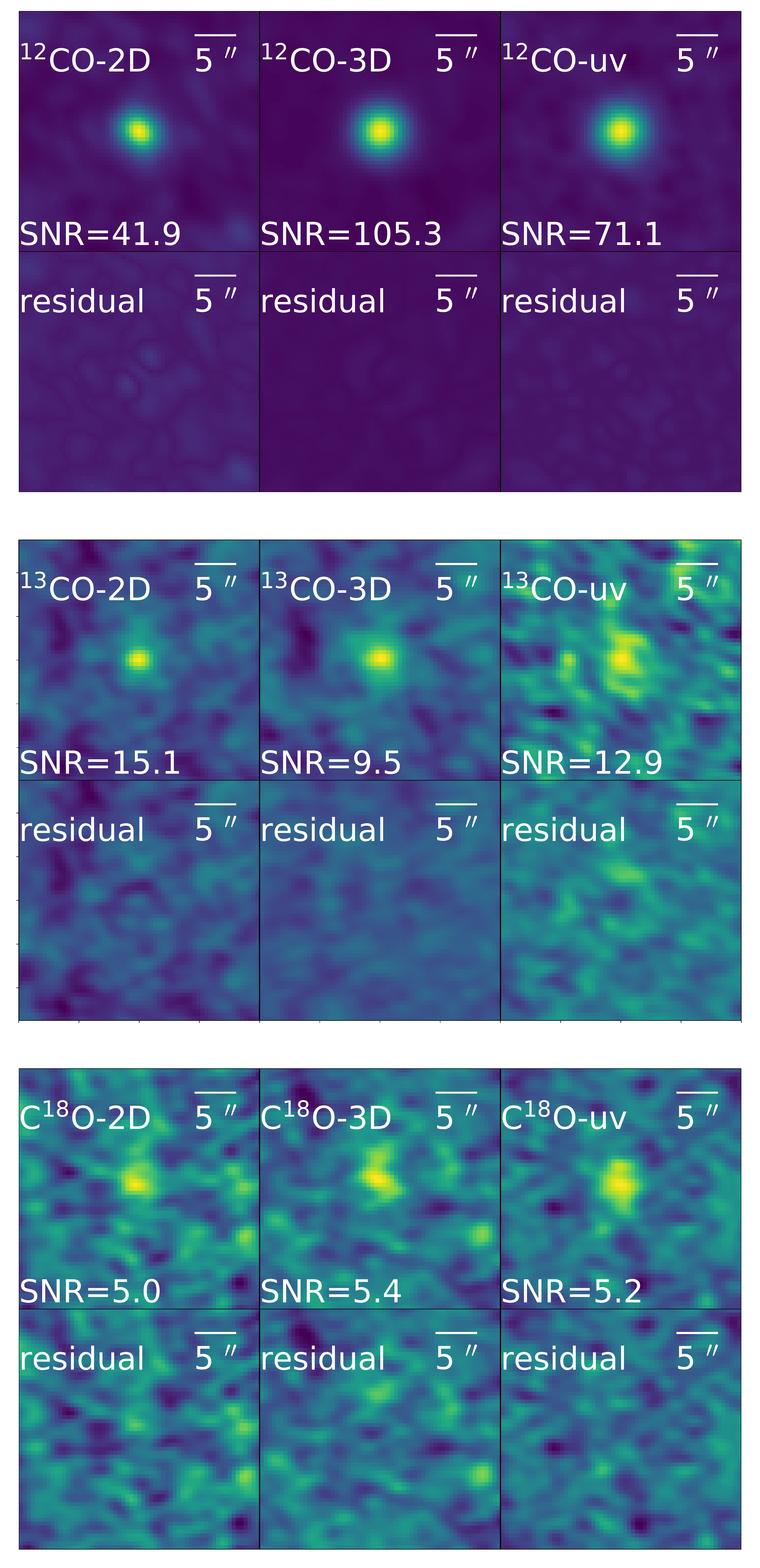}
\caption{Final composite 30"$\times$30" stamps stacks (top)and
residuals (bottom) from the corresponding flux modelling (see Section
\ref{sec:analysis}) for $^{12}$CO (upper panels), $^{13}$CO (middle
panels) and C$^{18}$O (lower panels) emission lines from 2D-moment-0
stack  (left column), 3D-image stack (middle column) and $uv$ stack
(right column).} \label{fig:UV_2D-3D-IMG_STACK} \end{figure}

In Fig.~\ref{fig:UV_2D-3D-IMG_STACK} we show 30"$\times$30" image 
stacks and residuals after point-source extraction for the three 
methods explored in this work. All three different approaches result 
in similar velocity integrated line flux densities within  the errors.
However,  we find that for a bright line like $^{12}$CO $uv$ stacks
shows a SNR $\sim$1.6$\times$ higher than that obtained from 2D
stacking method (see Table \ref{tbl:MthFlxs}).  This result is similar
to that found by \cite{Lindroos15} who reported  that continuum $uv$
stacking signal-to-noise ratio was up to 20\% higher than the
continuum image stacking. Nevertheless, we find that for  $^{12}$CO
3D-image stacking shows to be the method with the highest SNR, being
2.5 and 1.6 times higher than 2D-moment-0 stacking and  $uv$ stacking
SNRs respectively. On the other hand the 3D stacking  method shows the
lowest SNR for the faint lines like $^{13}$CO, while  the 2D-moment-0
and $uv$ stacking methods show similar SNRs. We note that,  all the
stacking methods applied on C$^{18}$O emission line show similar SNRs.
Even though $stacker$ was specially designed to stack $uv$ continuum
datasets and it has been successfully applied for $uv$ emission line
stacking  \citep{Fujimoto18,Fujimoto19,Fudamoto20,Carvajal20} using a
similar  procedure as described here, this is the first time that $uv$
and image  stacking methods for emission line observations are
directly compared. Driven by the previously available $^{12}$CO data
presented in  \cite{Villanueva17}, we decide to use the 2D (moment-0) 
approach to measure values, as this method yields the highest
$^{13}$CO  stacked SNRs that are straightforward to interpret and
simple to calculate

\begin{table} \caption{Signal-to-noise Ratio (SNR) detection 
for $^{12}$CO, $^{13}$CO and C$^{18}$O stacked line
emission, obtained from three different stacking methods explored in
this work 1) 2D-moment-0 stacking, 2) 3D-image stacking and 3) $uv$ stacking.}
\label{tbl:MthFlxs}
\begin{tabular}{|l|c|c|c|c}
\hline
 \multicolumn{1}{|c|}{SNR} &
 \multicolumn{1}{c|}{Moment-0} &
  \multicolumn{1}{c|}{$uv$-stacking} &
 \multicolumn{1}{c|}{3D-stacking} \\

\hline
\hline

$^{12}$CO & 41.9 & 71.1 & 105.3  \\
$^{13}$CO & 15.1 & 12.9 &  9.5   \\
C$^{18}$O &  5.0 &  5.2 &  5.4   \\

\hline\end{tabular}
\end{table}

\subsection{Luminosity measurements}\label{ssec:LumMes}

We compute the $^{12}$CO, $^{13}$CO and C$^{18}$O luminosities using
the velocity integrated line flux densities following
\citep{Solomon05}:

\begin{equation} L'_{\text{CO}}=3.25\times10^{7} S\Delta\text{v}
\nu_{\text{obs}}^{-2} {D_{\text{L}}}^{2}(1+z)^{-3}  \label{eq:S2L}
\end{equation}

\noindent where $L'$ $_{\text{CO}}$ is measured in
K\,km\,s$^{-1}$\,pc$^{2}$,  $S\Delta$v 
is the velocity integrated line flux density  in units of 
Jy\,km\,s$^{-1}$, $\nu_{\text{obs}}$ is the
observed frequency of  the emission line in GHz, $D_{\text{L}}$ is the
luminosity distance  in Mpc, and $z$ is the redshift.

Considering the fact that we are analysing average properties from
galaxies at different redshifts, special consideration should be taken
to convert to intrinsic luminosities. To determine the dispersion of
the stacked luminosity measurements we have used a Monte Carlo
simulation considering that the error  from the velocity integrated
line flux density measurements  is normally distributed, and at the
same time assume a random sampling for the redshift distribution of
the parent stacked sample. Repeating  this simulation, we get a
distribution of luminosities from which we  can then infer the
1$\sigma$ confidence intervals (CI) of our average luminosity
measurements. Since each galaxy  population has a different redshift
distribution, their CIs are independent from one population to
another. We note that the differences between intensity and luminosity
ratio measurements are negligible.  The luminosity ratio of any pair
of lines ($L'_{1}, L'_{2}$) comes from converting their fluxes into
luminosities following Equation \ref{eq:S2L}. Given that the redshifts
($z_{1},z_{2}$)  for both lines are the same, the redshift and
luminosity distance  dependencies vanish leading to Equation
\ref{eq:S2LR}.

\begin{equation}
\frac{L'_{1}}{L'_{2}} = 
\frac{S_{1}\Delta \text{v}_{1}\nu^{-2}_{1}D_{\text{L}_{1}}^{-2}(1+z_{1})^{-3}}
{S_{2}\Delta \text{v}_{2}\nu^{2}_{2}D_{\text{L}_{2}}^{2}(1+z_{2})^{-3}} = 
\frac{I_{1}\nu^{-2}_{1}}{I_{2}\nu^{-2}_{2}}
\label{eq:S2LR}
\end{equation}

In particular
$L'$\,($^{12}$CO\,)/$L'$\,($^{13}$CO)  =
0.91 $\times$ $I$\,($^{12}$CO\,)/$I$\,($^{13}$CO) and
$L'$\,($^{13}$CO\,)/$L'$\,(C$^{18}$O) = 0.99 $\times$
$I$\,($^{13}$CO\,)/$I$\,(C$^{18}$O), which enable us to make direct
comparisons between our results and different intensity and
luminosity ratios available in the literature.

\section{Results}\label{sec:Res} 
\subsection{The $L'$\,($^{12}$CO\,)/$L'$\,($^{13}$CO) ratio}
\label{sec:12CO13COR}  
The VALES survey provides a wide range of global galaxy properties such 
as stellar masses, star formation rates, morphologies, luminosities 
etc. In this section, we present the measured
$L'$\,($^{12}$CO\,)/$L'$\,($^{13}$CO) luminosity ratios to search for
possible dependencies on different global galaxy parameters. 

\subsubsection{Morphological and environmental dependence}
\label{ssec:R-Environ}

A morphological and environmental dependence of the 
$I$\,($^{12}$CO\,)/$I$\,($^{13}$CO) line intensity ratio has been
reported by previous studies. While merger systems show a higher
$I$\,($^{12}$CO\,)/$I$\,($^{13}$CO)  intensity ratio when compared
with normal spiral galaxies \citep{Casoli92b,Taniguchi98,Taniguchi99},
galaxies in dense environments show a lower
$I$\,($^{12}$CO\,)/$I$\,($^{13}$CO) intensity ratio \citep{Alatalo15}.
Initially, we explore the morphological and environmental
classification available for our sample, according to the most
prominent morphological features: Bulge (B), Disc (D),
Merger-Irregular (M), and (C) which denotes if the source has multiple
projected neighbouring systems (``companions''), as based on a visual
inspection presented by \cite{Villanueva17} to multi-wavelength
imaging from the GAMA survey.

\begin{figure}
\includegraphics[width=\columnwidth]{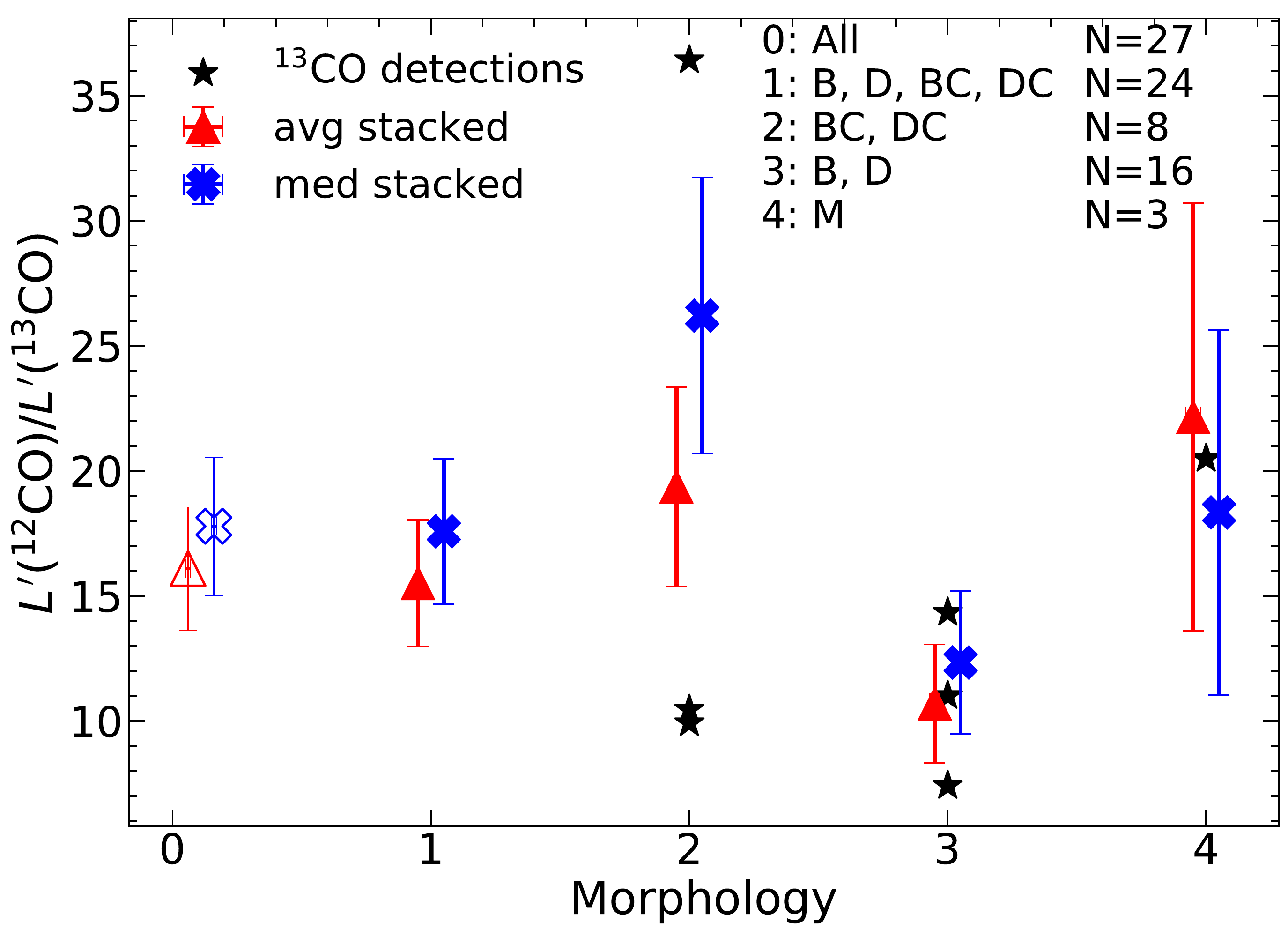}
\caption{Average (red triangles) and median (blue crosses) stacked 
$L'$\,($^{12}$CO\,)/$L'$\,($^{13}$CO)
line luminosity ratio values as a function of optical morphological
properties as presented in \protect\cite{Villanueva17}:
0) all galaxies (n=27) open symbols, 
1) all galaxies excluding mergers (B, D, BC, DC; n=24); 
2) bulge and disc dominated galaxies with projected companions (BC, DC; n=8);
3) bulge and disc dominated galaxies without any companion (B, D; n=16); and 
4) mergers (M; n=3). Error bars correspond to 1 $\sigma$ confidence intervals for 
average ratios.}
\label{fig:Mrp-L-12-13CO}
\end{figure}

We split our sample into 5 different subsets:  0) all galaxies; (n=27;
open symbols), 1) all galaxies excluding clear mergers (B, D, BC, DC;
n=24); 2) bulge and disc dominated galaxies with projected companions
(BC, DC; n=8); 3) bulge and disc dominated galaxies without any
companion (B, D; n=16); and 4) mergers (M; n=3) (see Figure
\ref{fig:Mrp-L-12-13CO}).\\

In Table \ref{tbl:MorphoRes} we present the measured average
$L'$\,($^{12}$CO\,)/$L'$\,($^{13}$CO) line luminosity ratio and
average  <SFR>, <SFE>, and <$L_{\text{IR}}$> values for the five
different morphological galaxy populations explored in this work. 
Using all of the 27 galaxies, we find an average
$L'$\,($^{12}$CO\,)/$L'$\,($^{13}$CO) line luminosity ratio of
16.1$\pm$2.5. This value is in agreement with the values for mergers
(12$\pm$3) and interacting early type galaxies (ETGs) (15$\pm$5)
reported by \cite{Alatalo15}, and to the ratio of nearby spirals, 
starburst and ETGs (excluding those belonging to the Virgo Cluster
used by \citealt{Alatalo15}) reported by \cite{Davis14} (12$\pm$1.0).
Mergers and galaxies with a visible companion tend to show higher
$L'$\,($^{12}$CO\,)/$L'$\,($^{13}$CO) line luminosity ratios. In
particular, mergers  show the highest <SFR>, <SFE>, and
<$L_{\text{IR}}$> average values  among the different morphological
classifications, and also show a $L'$\,($^{12}$CO\,)/$L'$\,($^{13}$CO)
line luminosity ratio 2 times higher  than that found in galaxies
without a companion. These findings, however, are at low significance 
 (mainly due to the low number statistics). Besides, these ratios are
in good agreement with the ratios reported by \cite{Alatalo15}, who
found that group galaxies present
$L'$\,($^{12}$CO\,)/$L'$\,($^{13}$CO) line luminosity ratios 2 times
higher than field galaxies.

\begin{table*}
\caption{Average $L'$\,($^{12}$CO\,)/$L'$\,($^{13}$CO) line luminosity ratio 
and average <SFR>, <SFE> and <$L_{\text{IR}}$> values
for different morphological classifications explored in this work
(\citealt{Villanueva17}; see \S~\ref{ssec:R-Environ} for more details).}
\label{tbl:MorphoRes} 
\begin{tabular}{|l|l|l|l|l|l|l|}
\hline
\multicolumn{1}{|c|}{ID} &
\multicolumn{1}{|c|}{Group} &
\multicolumn{1}{|c|}{N} & 
\multicolumn{1}{|c|}{$L'$($^{12}$CO)/$L'$($^{13}$CO)} & 
\multicolumn{1}{|c|}{<SFR>} &
\multicolumn{1}{|c|}{<SFE>} &
\multicolumn{1}{|c|}{<log[$L_{\text{IR}}$/$L_{\sun}$]>}
\\
\multicolumn{1}{|c|}{} &
\multicolumn{1}{|c|}{} &
\multicolumn{1}{|c|}{} & 
\multicolumn{1}{|c|}{} &
\multicolumn{1}{|c|}{M$_{\sun}$yr$^{-1}$ } &
\multicolumn{1}{|c|}{Gyr$^{-1}$} &
\multicolumn{1}{|c|}{}
\\
\hline
\hline
0 &All       &27&16.1 $\pm$ 2.5&14.9 $\pm$ 3.8&1.9 $\pm$ 0.5 & 10.9 $\pm$ 0.5 \\
1 &BC,DC,B,D &24&15.5 $\pm$ 2.5&12.7 $\pm$ 4.2&0.9 $\pm$ 0.1 & 10.8 $\pm$ 0.5 \\
2 &BC,DC     &16&19.4 $\pm$ 4.0&17.6 $\pm$ 5.9&1.0 $\pm$ 0.1 & 11.0 $\pm$ 0.5 \\
3 &B,D       & 8&10.7 $\pm$ 2.4& 3.4 $\pm$ 0.5&0.8 $\pm$ 0.1 & 10.5 $\pm$ 0.2 \\
4 &M         & 3&22.1 $\pm$ 8.6& 34  $\pm$ 6.9&9.6 $\pm$ 0.8 & 11.5 $\pm$ 0.2 \\
\end{tabular}
\end{table*}

\subsubsection{The star-formation activity}\label{ssec:R-SSFR}

\begin{figure*}
\includegraphics[width=\textwidth]{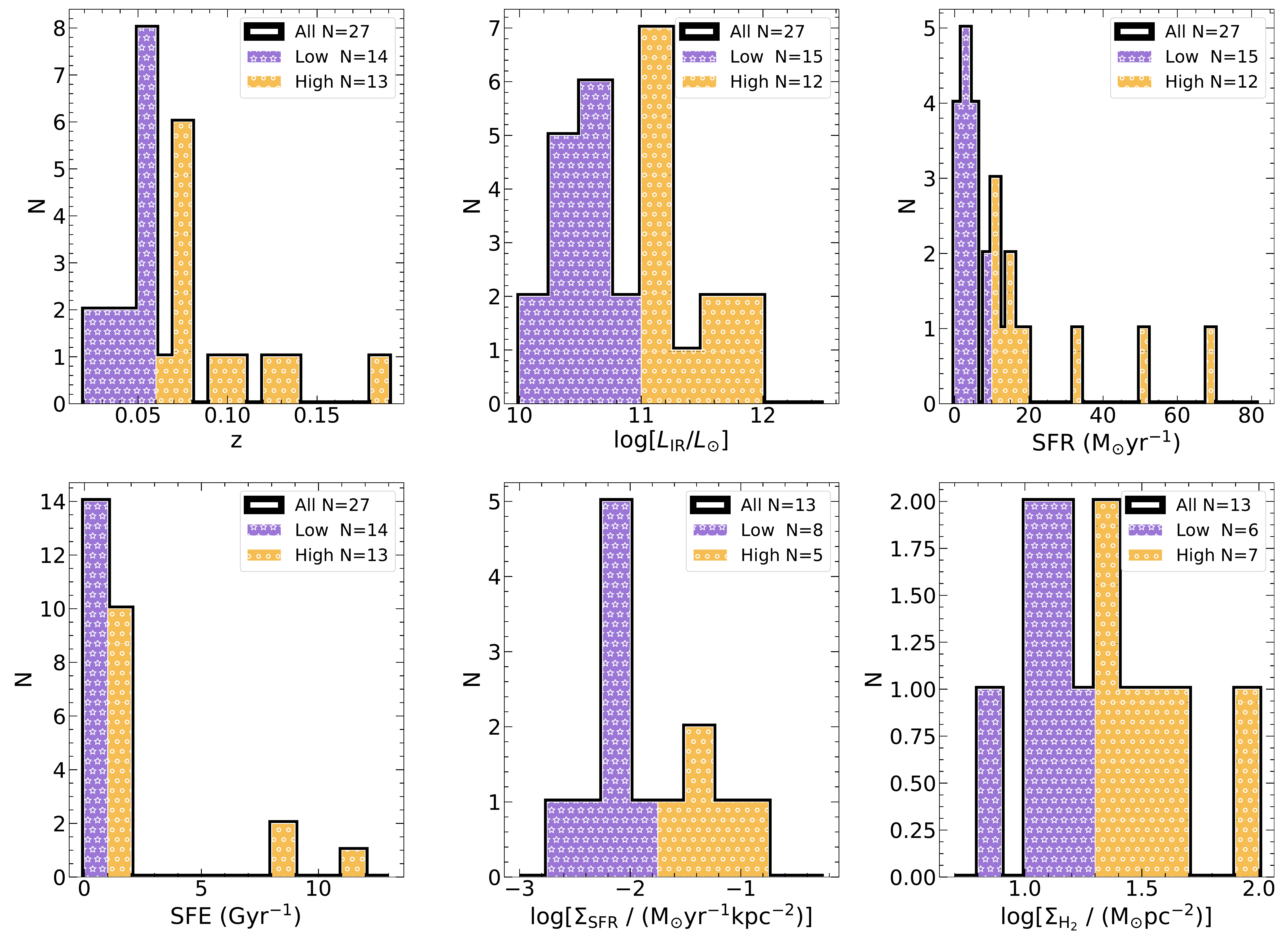}
\caption{Redshift, $L_{\text{IR}}$, SFR, SFE, $\Sigma_{\text{SFR}}$
and $\Sigma_{\text{H}_{2}}$ distributions of the galaxies used in the
$^{12}$CO and $^{13}$CO stacking analysis (solid black line),  split
by low (purple bars hatched with stars) and high  (yellow bars hatched
with circles) values.} \label{fig:Histogram} \end{figure*}

\begin{figure}
\includegraphics[width=\columnwidth]{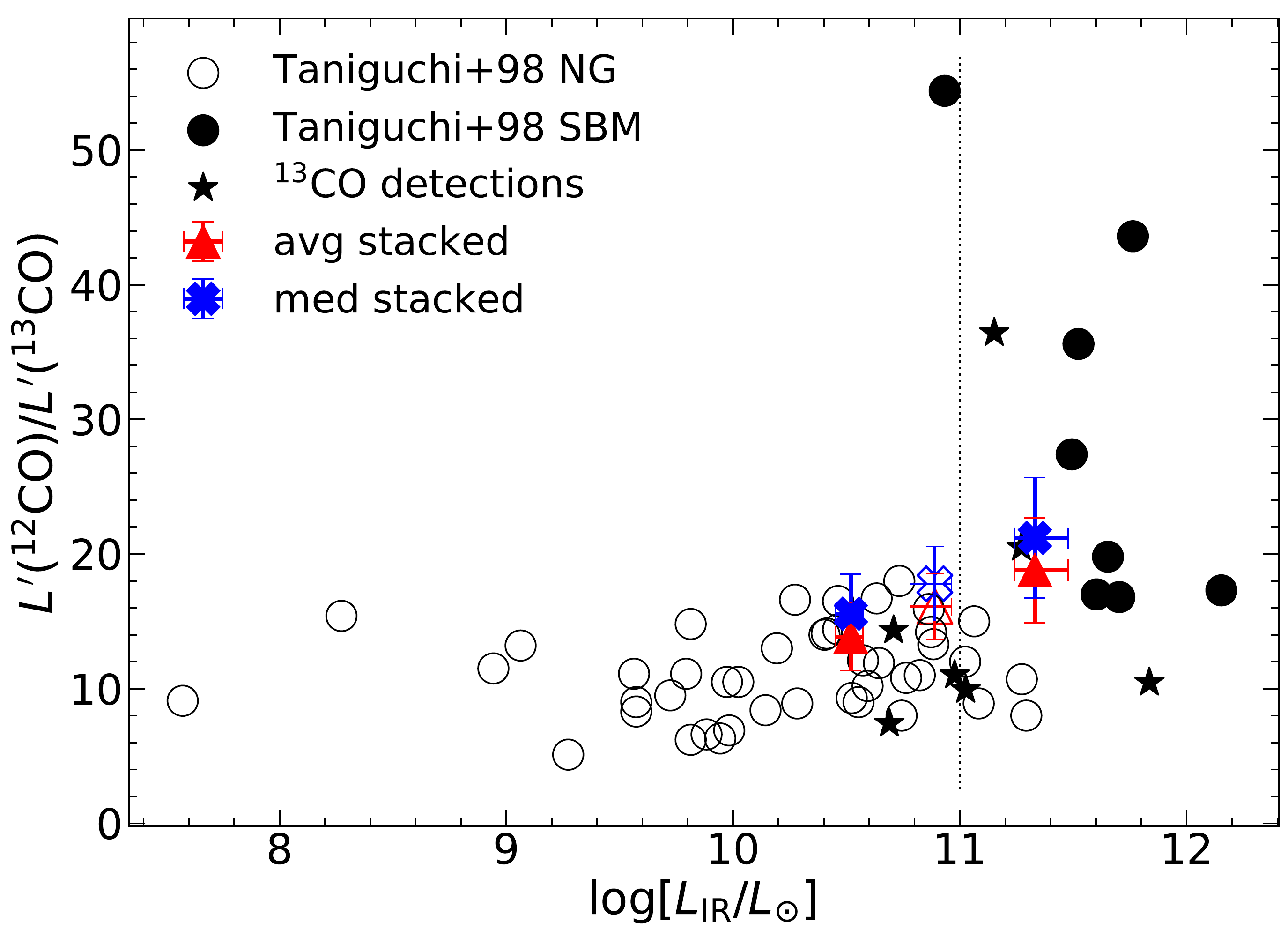}
\caption{Average (red triangles) and median (blue crosses) stacked 
$L'$\,($^{12}$CO\,)/$L'$\,($^{13}$CO) line luminosity ratio for low
and high $L_{\text{IR}}$ subsets (triangles). A dashed line indicates
the boundary between low and high $L_{\text{IR}}$ populations. As
reference average and median line  luminosity ratio considering all
galaxies in open symbols are included.  Error bars correspond to
1$\sigma$ confidence intervals for average or  median values based
Monte-Carlo simulations. Individual $^{13}$CO  detections (stars) and
$L'$\,($^{12}$CO\,)/$L'$\,($^{13}$CO) line  luminosity ratios of
normal galaxies (NG open circles) and starburst mergers (SBM filled
circles) scaled by a 1.75 factor to convert from far-IR to IR 
luminosities (see Appendix E from \citealt{HerreraCamus15}) from
\protect\cite{Taniguchi98}are also included.}
\label{fig:LFIR-L-12-13CO} \end{figure}

\begin{figure*}
\includegraphics[width=\textwidth]{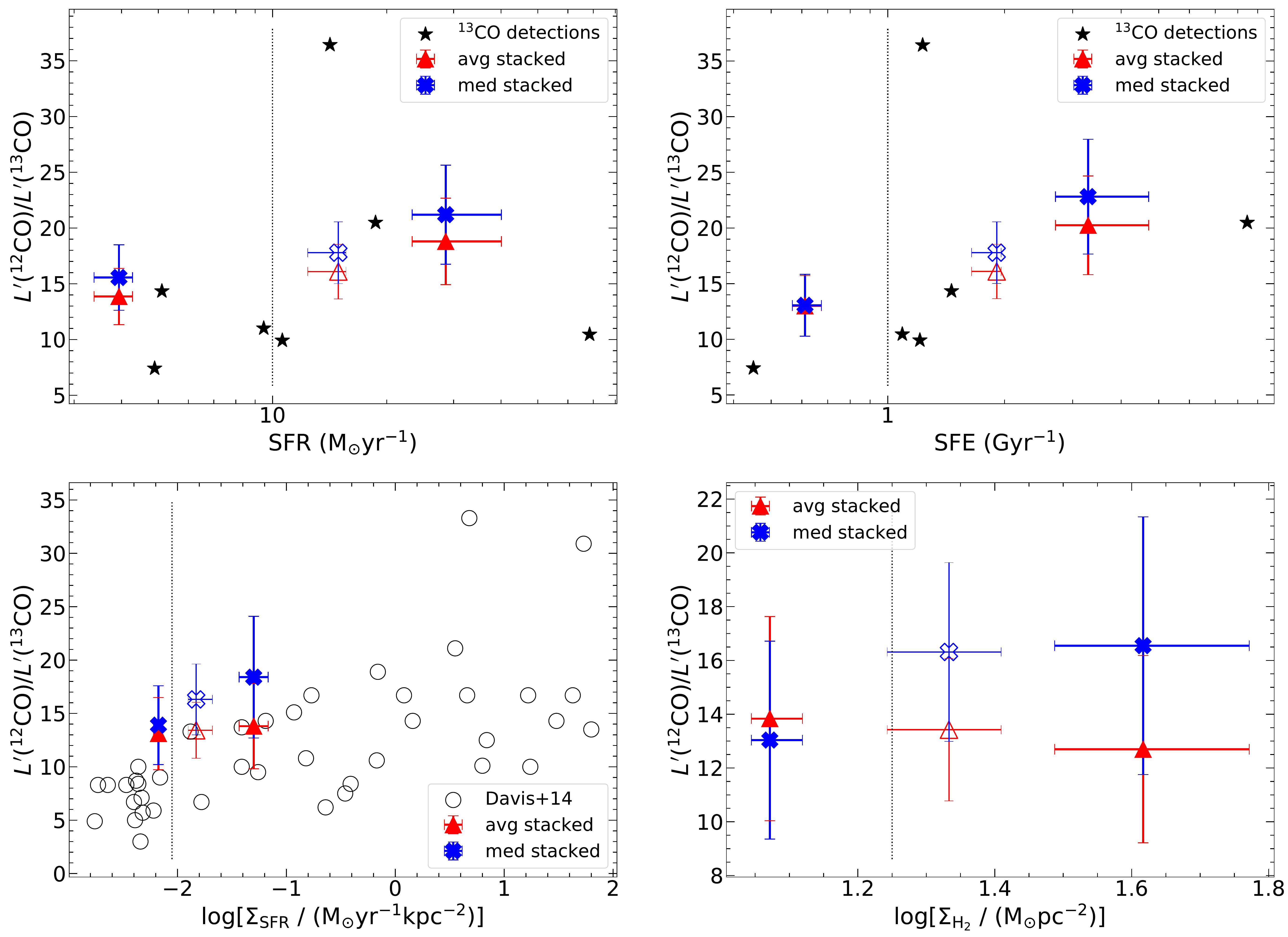}
\caption{Average (red triangles) and median (blue crosses) stacked
$L'$\,($^{12}$CO\,)/$L'$\,($^{13}$CO) line luminosity ratio for low
and high SFR (upper left), SFE (upper right), $\Sigma_{\text{SFR}}$
(lower left) and $\Sigma_{\text{H}_{2}}$ (lower right) subsets
(triangles). As reference average and median line luminosity ratio
considering all galaxies in open symbols and  dashed lines indicating
the  boundary between low and high SFR, SFE, $\Sigma_{\text{SFR}}$ and
 $\Sigma_{\text{H}_{2}}$ populations are included. Error bars
correspond to 1$\sigma$ confidence intervals for average or median
values based on Monte-Carlo simulations. Individual $^{13}$CO
detections (stars) and $I$\,($^{12}$CO\,)/$I$\,($^{13}$CO) line
intensity ratio of normal galaxies (open circles) from
\protect\cite{Davis14} are also included.} \label{fig:MLT-L-12-13CO}
\end{figure*}

\cite{Villanueva17} derived various global galaxy properties,
including star formation rate (SFR), star formation efficiency (SFE),
molecular gas surface density ($\Sigma_{\text{H}_{2}}$), star
formation rate surface density ($\Sigma_{\text{SFR}}$), stellar mass
(M$_{\star}$), gas depletion time ($\tau$), and projected size
(R$_{\text{FWHM}}$).  The total IR luminosity was obtained as
described in \cite{Ibar15} by integrating the best-fitting SED between
8 and 1000 $\mu$m  using photometry from IRAS, Wide-field Infrared
Survey Explorer (WISE), and {\it Herschel}. The star formation rate
was estimated  following SFR (M$_\odot$yr$^{-1}$) = 10$^{-10}\times$
$L_{\text{IR}}$  assuming a \citet{Chabrier03} IMF. The molecular gas
mass was computed using $L'_{\text{CO}}$ and assuming an
$\alpha_{\text{CO}}$ conversion  factor dependent on the morphological
classification  ($\alpha_{\text{CO}}=4.6$ K km s$^{-1}$ pc$^{2}$ for B
and D dominated galaxies and $\alpha_{\text{CO}}=0.8$ K km s$^{-1}$
pc$^{2}$ for mergers/interacting galaxies). The SFR and
M$_{\text{H}_{2}}$ surface densities were estimated by dividing the
measured values by the area of a two-sided disc (2$\pi
R^{2}_{\text{FWHM}}$),   where $R_{\text{FWHM}}$ is the  deconvolved
FWHM along the semimajor axis obtained through fitting elliptical
gaussian profiles  to the $^{12}$CO ($1-0$) moment-0 maps using the 
CASA task {\sc imfit}. We consider the CO  emission to be spatially
resolved if the fitted semimajor axis is at least $\sqrt{2}$ times
larger than the semimajor axis of the  synthesized beam. A more
detailed discussion about the computations of these parameters can be
found in \citet{Villanueva17}. With these in hand, we looked for
possible dependencies of $L'$\,($^{12}$CO\,)/$L'$\,($^{13}$CO)
luminosity line ratio on these global galaxy properties by splitting 
our sample in two bins for each parameter. Figure \ref{fig:Histogram}
shows the redshift, $L_{\text{IR}}$, SFR, SFE, $\Sigma_{\text{SFR}}$
and $\Sigma_{\text{H}_{2}}$ distributions, split by low  and high
values. We find that the most significant trends for the
$L'$\,($^{12}$CO\,)/$L'$\,($^{13}$CO) ratios are with $L_{\text{IR}}$,
SFR, and SFE (see Figures \ref{fig:LFIR-L-12-13CO} and
\ref{fig:MLT-L-12-13CO}). Table  \ref{tbl:SplitSamplVal1} shows the
average $L'$\,($^{12}$CO\,)/$L'$\,($^{13}$CO) line  luminosity ratios
for low and high $L_{\text{IR}}$, SFR, SFE, $\Sigma_{\text{SFR}}$ and
$\Sigma_{\text{H}_{2}}$ populations.

\begin{table*}
\caption{Average $L'$\,($^{12}$CO\,)/$L'$\,($^{13}$CO) line luminosity ratio 
for our sample of galaxies split by high and low  $L_{\text{IR}}$, SFR, SFE,
$\Sigma_{\text{SFR}}$ and $\Sigma_{\text{H}_{2}}$ values. 
Column 1: Parameter of interest; 
Column 2: Range explored;
Column 3: Number of galaxies in the explored range;
Column 4: Average value for parameter of interest;
Column 5: The average stacked $L'$\,($^{12}$CO\,)/$L'$\,($^{13}$CO) line 
luminosity ratio;
Column 6 and 7: Student's t-test statistical reports (t, p) to asses the
probability (p) that the null hypothesis 
($L'$\,($^{12}$CO\,)/$L'$\,($^{13}$CO) line luminosity 
ratio variations between the low and high SFR populations is not 
statistically significant), is true.}
\label{tbl:SplitSamplVal1}
\begin{tabular}{|l|l|l|l|l|l|l|}
\hline
\multicolumn{1}{|c|}{\multirow{2}{*}{Parameter}} &
\multicolumn{1}{|c|}{\multirow{2}{*}{Range}} &
\multicolumn{1}{|c|}{\multirow{2}{*}{N}} & 
\multicolumn{1}{|c|}{\multirow{2}{*}{Average}} &
\multicolumn{1}{|c|}{\multirow{2}{*}{$L'$($^{12}$CO)/$L'$($^{13}$CO)}} &
\multicolumn{2}{|c|}{t-test} \\
& & & & &
\multicolumn{1}{|c|}{t} &
\multicolumn{1}{|c|}{p} \\

\multicolumn{1}{|c|}{1} &
\multicolumn{1}{|c|}{2} &
\multicolumn{1}{|c|}{3} &
\multicolumn{1}{|c|}{4} &
\multicolumn{1}{|c|}{5} &
\multicolumn{1}{|c|}{6} &
\multicolumn{1}{|c|}{7}
\\
\hline
\hline
\\
\multirow{2}{*}{log[$L_{\text{IR}}$/L$_{\odot}$]} &
  [10.1 - 10.9]    & 14 & 10.5 $\pm$ 0.1 & 13.8 $\pm$ 2.4 & 
  \multirow{2}{*}{-3.9} &
  \multirow{2}{*}{0.0006} \\
 & [11.0 - 11.9]   & 13 & 11.3 $\pm$ 0.1 & 18.7 $\pm$ 3.9\\
\hline 
\multirow{2}{*}{SFR (M$_{\odot}${yr}$^{-1}$)} &
  [1.8 - 9.5]      & 15 &  3.9 $\pm$ 0.6 & 13.3 $\pm$ 2.4 & 
  \multirow{2}{*}{-4.3} &
  \multirow{2}{*}{0.0002}  \\
 & [10.3 - 83.4]   & 12 & 28.6 $\pm$ 5.3 & 18.7 $\pm$ 3.8\\
\hline
\multirow{2}{*}{SFE (Gyr$^{-1}$)} &
  [0.4 - 0.9]    & 14 & 0.6  $\pm$ 0.1 & 12.9 $\pm$ 2.7 & 
  \multirow{2}{*}{-4.7} &
  \multirow{2}{*}{7E-05} \\
 & [1  - 11.7]   & 13 & 3.3  $\pm$ 0.6 & 19.4 $\pm$ 4.2\\
\hline
\multirow{3}{*}{log[$\Sigma_{\text{SFR}}$ \, / (M$_{\odot}${yr}$^{-1}$kpc$^{-2}$)]} &
  [-2.6 - -1]     & 13 & -1.8 $\pm$ 0.1 & 12.5 $\pm$ 2.4\\
 & [-2.6 - -2]    &  8 & -2.2 $\pm$ 0.1 & 12.2 $\pm$ 3.1 & 
\multirow{2}{*}{-0.05} &
\multirow{2}{*}{0.9} \\
 & [-2.1 - -1]    &  5 & -1.3 $\pm$ 0.1 & 12.3 $\pm$ 3.5\\
\hline
\multirow{3}{*}{log[$\Sigma_{\text{H}_{2}}$ \, / (M$_{\odot}$pc$^{-2}$)]} &
  [0.8 - 2.0]     & 13 & 1.3 $\pm$ 0.1 & 12.5 $\pm$ 2.5\\
 & [0.8 - 1.2]    & 6  & 1.0 $\pm$ 0.1 & 11.8 $\pm$ 3.2 & 
\multirow{2}{*}{-0.5} &
\multirow{2}{*}{0.6} \\
 & [1.3 - 2.0]    & 7  & 1.6 $\pm$ 0.2 & 12.8 $\pm$ 3.5\\
\end{tabular}
\end{table*}

Figure \ref{fig:LFIR-L-12-13CO} shows a trend of
$L'$\,($^{12}$CO\,)/$L'$\,($^{13}$CO) line luminosity ratio with
$L_{\text{IR}}$ similar to that shown by \cite{Taniguchi98} who used
$^{12}$CO and $^{13}$CO line data taken from the literature for 61
nearby galaxies, including eight luminous starburst galaxies. They
found a correlation between L$_{\text{FIR}}$ and 
$L'$\,($^{12}$CO\,)/$L'$\,($^{13}$CO) line ratio, where starburst
galaxies with high infrared luminosities show higher 
($L'$\,($^{12}$CO\,)/$L'$\,($^{13}$CO)\,$\geq$\,20) line ratios
compared with normal galaxies. They dismissed physical gas properties
such as density, temperature, or velocity gradients as responsible for
the observed high   $L'$\,($^{12}$CO\,)/$L'$\,($^{13}$CO) abundance
ratio in starburst galaxies  and conclude that the only possible
mechanism behind the high $^{12}$CO/$^{13}$CO abundance ratio in
starburst  galaxies is an underabundance of $^{13}$CO with respect to
$^{12}$CO. Our stacks (see Figure \ref{fig:LFIR-L-12-13CO}) show a
trend in which the high $L_{\text{IR}}$ sample falls in the starburst
region, while the  low $L_{\text{IR}}$ sample shows an average
$L'$\,($^{12}$CO\,)/$L'$\,($^{13}$CO) line luminosity ratio similar to
that found in normal galaxies.

To test the significance of the
$L'$\,($^{12}$CO\,)/$L'$\,($^{13}$CO)-$L_{\text{IR}}$ variations we
applied a Student's t-test to determine the probability  that the
$L'$\,($^{12}$CO\,)/$L'$\,($^{13}$CO) line luminosity  ratio
variations between the low and high $L_{\text{IR}}$   populations are
not statistically significant.  A large p-value indicates that the
differences between the two sample  means are not statistically
significant, while a small one suggests  that the differences between
the two sample means are significant.  Based on this test we find that
the differences in the 
$L'$\,($^{12}$CO\,)/$L'$\,($^{13}$CO)-$L_{\text{IR}}$ variations  are
statistically significant (see Table \ref{tbl:SplitSamplVal1}).  We
then implemented a Spearman rank test to investigate   a possible
$L'$\,($^{12}$CO\,)/$L'$\,($^{13}$CO)-$L_{\text{IR}}$ correlation, 
where p-values report the probability of the lack of correlation
between the two samples. A large p-value indicates that there is no
significant correlation,  while a small one suggests a significant
correlation. The Spearman rank  test was computed considering six
galaxies (excluding HATLASJ090949.6  identified as an outlier see
Section \ref{ssec:Indv}) with individual  $^{13}$CO detections, for
which we could compute individual 
$L'$\,($^{12}$CO\,)/$L'$\,($^{13}$CO) luminosity ratios. The Spearman
rank test does not provide  evidence (see Table \ref{tbl:SpearmanTest}
first row) supporting a significant correlation between
$L'$\,($^{12}$CO\,)/$L'$\,($^{13}$CO) and  $L_{\text{IR}}$. This might
be due to  the reduced number (6) of galaxies with individual
$^{13}$CO detections covering a small range of $L_{\rm IR}$. However,
if we also consider starburst galaxies and normal galaxies  covering a
wider range of $L_{\rm FIR}$ as reported by
\protect\cite{Taniguchi98},  we find (see Table \ref{tbl:SpearmanTest}
second row) evidence supporting a moderate 
$L'$\,($^{12}$CO\,)/$L'$\,($^{13}$CO)-$L_{\text{IR}}$ correlation.

\begin{table}\caption{Spearman correlation test statistic 
($\rho$, p), to asses the null hypothesis (no significant 
correlation between $L'$\,($^{12}$CO\,)/$L'$\,($^{13}$CO) and 
$L_{\text{IR}}$) being true, for 1) 6 galaxies with $^{13}$CO individual 
detections for which we could compute individual $L'$\,($^{12}$CO\,)/$L'$\,($^{13}$CO) 
luminosity ratios and 2) 6 galaxies with $^{13}$CO individual
detections plus 61 starburst and normal galaxies reported by \protect\cite{Taniguchi98}.}
\label{tbl:SpearmanTest} 
\begin{tabular}{|l|c|c|c|c}
\hline
 \multicolumn{1}{|c|}{Sample} &
 \multicolumn{1}{c|}{N} &
  \multicolumn{1}{c|}{$\rho$} &
 \multicolumn{1}{c|}{p} \\
\hline
\hline
1) Galaxies with $^{13}$CO detections &  6 & 0.71 & 0.14  \\
2) + Starburst and Normal galaxies   & 67 & 0.55 & 4E-6 \\
\hline\end{tabular}
\end{table}

Figure \ref{fig:MLT-L-12-13CO} (upper panels) shows 
$L'$\,($^{12}$CO\,)/$L'$\,($^{13}$CO) line luminosity ratio trends
with SFR and SFE. Considering that SFRs are derived from far-IR
luminosities, an expected  trend in SFR is also identified (see
Fig.~\ref{fig:LFIR-L-12-13CO}).  We also find that the higher the SFE
(SFR/M$_{\text{H}_{2}}$), the higher the
$L'$\,($^{12}$CO\,)/$L'$\,($^{13}$CO) line luminosity ratio, following
an expected similar trend as with SFR.  By looking at the stacked
signals, we find significant variations of the
$L'$\,($^{12}$CO\,)/$L'$\,($^{13}$CO) line luminosity ratio when we
split our sample by low and high SFR values (see
Table~\ref{tbl:StatSplitHighLow}).  We notice that galaxies with high
SFR not only show high  $L'$\,($^{12}$CO\,)/$L'$\,($^{13}$CO)
luminosity ratio, but also show relatively high reservoirs of
molecular gas.  Table~\ref{tbl:StatSplitHighLow} shows the average
values of  redshift, molecular gas mass (M$_{\text{H}_{2}}$),
molecular gas mass to  stellar mass ratio
(M$_{\text{H}_{2}}$/M$_{\star}$), and molecular  gas fraction ($f_{\,
\text{H}_{2}}$ = M$_{\text{H}_{2}}$ /
(M$_{\text{H}_{2}}$+M$_{\star}$)) for our galaxy sample after
splitting  it in low and high SFR values. To estimate the significance
of the  observed differences between the average properties 
(M$_{\text{H}_{2}}$, M$_{\text{H}_{2}}$/M$_{\star}$, $f_{\,
\text{H}_{2}}$)  in the low and high SFR populations, we applied a
Kolmogorov-Smirnov (KS)  test to compute the probability that low and
high  ($z$, $M_{\text{H}_{2}}$/$M_{\star}$,
$M_{\text{H}_{2}}$/$M_{\star}$ and  $f_{\, \text{H}_{2}}$)
distributions were drawn from from the same parent population. A large
p-value indicates  that the distributions are identical, while a small
one suggests that  the distributions are different.  In all cases we
find evidence to reject the null hypothesis that the observed
properties in the low and high SFR populations were  drawn from the
same parent population (see Table ~\ref{tbl:StatSplitHighLow}).

\begin{table}
\caption{Average values of various galaxy parameters (Column 1):
redshift, molecular gas mass (M$_{\text{H}_{2}}$), molecular gas mass to 
stellar mass ratio (M$_{\text{H}_{2}}$/M$_{\star}$), and molecular 
gas fraction ($f_{\, \text{H}_{2}}$ = M$_{\text{H}_{2}}$
/ (M$_{\text{H}_{2}}$+M$_{\star}$))
for the sample after splitting it in low (Column 2) and 
high (Columns 3) SFR values.
Columns 4 and 5 contain the Kolmogorov-Smirnov statistical 
reports (D, p) to asses the probability 
that low and high ($z$, 
$M_{\text{H}_{2}}$/$M_{\star}$ and $f_{\, \text{H}_{2}}$) 
were drawn from populations with identical 
distributions.} 
\label{tbl:StatSplitHighLow} 
\begin{tabular}{|l|r|r|r|r|}
\hline
  & \multicolumn{2}{c}{SFR [M$_{\odot}$yr$^{-1}$]}\\
\hline
 \multicolumn{1}{|c|}{Parameter} &
 \multicolumn{1}{c|}{low} &
 \multicolumn{1}{c|}{high} &
 \multicolumn{2}{c|}{KS}\\
 \multicolumn{1}{|c|}{} &
 \multicolumn{1}{c|}{N = 15} &
 \multicolumn{1}{c|}{N = 12} &
 \multicolumn{1}{|c|}{\multirow{2}{*}{D}} &
 \multicolumn{1}{|c|}{\multirow{2}{*}{p}}\\
 \multicolumn{1}{|c|}{} &
 \multicolumn{1}{c|}{[1.8-9.5]} &
 \multicolumn{1}{c|}{[10.3-83.4]} \\
\multicolumn{1}{|c|}{1} &
\multicolumn{1}{|c|}{2} &
\multicolumn{1}{|c|}{3} &
\multicolumn{1}{|c|}{4} &
\multicolumn{1}{|c|}{5} \\
\hline
\hline
 <z>                                   & 0.04 $\pm$ 0.01 & 0.10 $\pm$ 0.04 
                                       & 0.8 &0.0001\\
 <log[M$_{\text{H}_{2}}$/M$_{\odot}$]> & 8.9  $\pm$ 0.6  & 10.1 $\pm$ 0.1  
                                       & 0.5 &0.03\\
 <M$_{\text{H}_{2}}$/M$_{\star}$>      & 0.17 $\pm$ 0.04 & 0.53 $\pm$ 0.18 
                                       & 0.6 &0.02\\
 <$f_{\, \text{H}_{2}}$>               & 0.14 $\pm$ 0.06 & 0.28 $\pm$ 0.15 
                                       & 0.55 &0.02\\
\end{tabular}
\end{table}

We perform a similar analysis to that described by \cite{Davis14} that
reported a positive  correlation of the
$I$\,($^{12}$CO\,)/$I$\,($^{13}$CO) line intensity ratio with
$\Sigma_{\text{SFR}}$ and $\Sigma_{\text{H}_{2}}$. In our case  only
13 galaxies are known to be resolved in $^{12}$CO, enabling us to
derive $\Sigma_{\text{SFR}}$ and   $\Sigma_{\text{H}_{2}}$. In
Figure~\ref{fig:MLT-L-12-13CO} (lower panels)  we show the
$L'$\,($^{12}$CO\,)/$L'$\,($^{13}$CO) line luminosity ratio as a
function of $\Sigma_{\text{SFR}}$ split in high 
(10$^{-2.1}$--10$^{-1}$\,M$_{\odot}$yr$^{-1}$\,kpc$^{-2}$) and low 
(10$^{-2.6}$--10$^{-2}$\,M$_{\odot}$yr$^{-1}$\,kpc$^{-2}$) values.
Individual $^{13}$CO galaxy detections from \cite{Davis14} are also
over-plotted. Concerning the dependencies of 
$L'$\,($^{12}$CO\,)/$L'$\,($^{13}$CO) as a function of
$\Sigma_{\text{H}_{2}}$, when we split by low 
(10$^{0.8}$--10$^{1.3}$\,M$_{\odot}$\,pc$^{-2}$) and high 
(10$^{1.3}$--10$^{-2}$\,M$_{\odot}$\,pc$^{-2}$) values,   a moderate
trend is found.

We applied a Student's t-test to evaluate the probability that the
null  hypothesis (i.e. that the $L'$\,($^{12}$CO\,)/$L'$\,($^{13}$CO)
line luminosity  ratio variations between the low and high  SFR, SFE, 
$\Sigma_{\text{SFR}}$ and $\Sigma_{\text{H}_{2}}$ populations are not
statistically significant), is true. We find supporting evidence that 
$L'$\,($^{12}$CO\,)/$L'$\,($^{13}$CO) variations between low and high
SFR and SFE populations are statistically significant.  On the other
hand we do not find evidence that supports   that
$L'$\,($^{12}$CO\,)/$L'$\,($^{13}$CO) variations between low and high
$\Sigma_{\text{SFR}}$, $\Sigma_{\text{H}_{2}}$ populations are not
statistically significant (see Table  \ref{tbl:SplitSamplVal1}). The
lack of a significant difference of
$L'$\,($^{12}$CO\,)/$L'$\,($^{13}$CO) with both  $\Sigma_{\text{SFR}}$
and $\Sigma_{\text{H}_{2}}$  shown in the lower panels of
Figure~\ref{fig:MLT-L-12-13CO}  is most probably due to  the reduced
number of galaxies used for $\Sigma_{\text{SFR}}$  and
$\Sigma_{\text{H}_{2}}$ stacks and the relatively small range of
surface densities explored in this work.  Finally, we note that we did
not find any significant variations of 
$L'$\,($^{12}$CO\,)/$L'$\,($^{13}$CO) with redshift and stellar mass,
as with $\Sigma_{\text{H}_{2}}$. This might be caused by the 
relatively small range of redshift ([0.025-0.195]) and stellar masses
(log(M/M$_{\odot}$)=9.8-10.9) explored in this work.

\subsubsection{$^{13}$CO Individual detections}\label{ssec:Indv}

We have included the $L'$\,($^{12}$CO\,)/$L'$\,($^{13}$CO) line
luminosity ratios of the individual $^{13}$CO detections (see Table
\ref{tbl:13COIndvDet}) in Figures \ref{fig:Mrp-L-12-13CO},
\ref{fig:LFIR-L-12-13CO}, and \ref{fig:MLT-L-12-13CO}. Galaxy
J085748.0+004641  is classified as merger (M) while galaxy
J083831.8+000044 is  classified as a galaxy with a projected companion
(DBC), they both show high $L'$\,($^{12}$CO\,)/$L'$\,($^{13}$CO) line
luminosity ratios, in good agreement with previous findings. On the
other hand, galaxies J085346.4+001252, J084139.6+015346, 
J084350.8+005534 and J090633.6+001526 are classified as galaxies that
do not show any apparent projected companion and present low
$L'$\,($^{12}$CO\,)/$L'$\,($^{13}$CO) line luminosity ratios, in
agreement with those expected from galaxies with low $L_{\text{IR}}$
values. Finally, galaxy J090949.6+014847 seems to be a peculiar galaxy
showing a low $L'$\,($^{12}$CO\,)/$L'$\,($^{13}$CO) line luminosity
ratio but with low SFE\,$=$\,1\,Gyr$^{-1}$ and  high
$L_{\text{IR}}$\,$=$\,10$^{12}$\,L$_{\odot}$ values. Thus, we identify
J090949.6+014847 as an outlier and exclude it from the  Spearman rank
tests in the previous analyses. Stacking results are robust against
the removal of this peculiar source from the sample.

\subsubsection{110.201 GHz stacked continuum
emission.}\label{ssec:ContStacks} As discussed in section
\ref{ssec:13COObs} we do not detect continuum emission at $\sim$110GHz
above $5\sigma$ significance down to a rms\  noise of
4$\mu$Jy\,beam$^{-1}$ in any of the 27 galaxies of our sample. 
However, we could detect a high signal-to-noise ratio (SNR=13) 
emission after stacking the individual continuum emission of our 27 
galaxy sample coming from the $^{13}$CO datasets. As with the
$^{12}$CO, $^{13}$CO and C$^{18}$O stacks, we split our sample by low
and high SFR populations (see Table \ref{tbl:StatSplitHighLowCont}).
Similar to what we found for the $L'$\,($^{12}$CO\,)/$L'$\,($^{13}$CO)
ratio, galaxies with higher SFR show the higher continuum emission
compared with galaxies with low SFR. The detected continuum emission
does not show any discrepancy with the expected average continuum
emission at  $\sim$110 GHz extrapolated from the spectral energy
distributions presented in \cite{Villanueva17}. However further
analyses (beyond the scope of this paper) at these frequencies are 
needed to uncover the origin of the continuum emission  (\textit{e.g.
free-free, dust, ionized gas emission})  contributing to the SEDs of
these galaxies and its relation with  their star formation activity.

\begin{table}
\caption{$^{13}$CO stacked continuum emission split by 
low and high SFR populations.} 
\label{tbl:StatSplitHighLowCont} 
\begin{tabular}{|l|r|r|r|r|}
\hline
  & \multicolumn{2}{c}{SFR [M$_{\odot}$yr$^{-1}$]}\\
\hline
 \multicolumn{1}{|c|}{Parameter} &
\multicolumn{1}{c|}{All} &
 \multicolumn{1}{c|}{low} &
 \multicolumn{1}{c|}{high}\\
 \multicolumn{1}{|c|}{} &
 \multicolumn{1}{c|}{N = 27} &
 \multicolumn{1}{c|}{N = 15} &
 \multicolumn{1}{c|}{N = 12} & \\
 \multicolumn{1}{|c|}{} & &
 \multicolumn{1}{c|}{[1.8-9.5]} &
 \multicolumn{1}{c|}{[10.3-83.4]} \\
\hline
\hline
 SNR                         & 13        & 6             &15\\
 S$_{^{13}\text{CO}} \Delta$v [$\mu$Jy\,km\,s$^{-1}$]    & 82 $\pm$ 6 & 64 $\pm$ 10 & 110 $\pm$ 7\\
\end{tabular}
\end{table}

\subsection{$L'$\,($^{13}$CO\,)/$L'$\,(C$^{18}$O) correlations.}\label{ssec:C18O}

\begin{table*}
\caption{Average $L'$\,($^{12}$CO\,)/$L'$\,($^{13}$CO) and 
$L'$\,($^{13}$CO\,)/$L'$\,(C$^{18}$O) line luminosity ratios for the sample
of 24 VALES galaxies with C$^{18}$O observations 
split by low and high $L_{\text{IR}}$, SFR, SFE, values (see Figs.
~\ref{fig:C18OStackResultsTrends1213} and ~\ref{fig:C18OStackResultsTrends}). 
Column 1: Parameter of interest; 
Column 2: Range explored;
Column 3: Number of galaxies in the explored range;
Column 4: Average value for parameter of interest;
Columns 5 and 8: The average stacked $L'$\,($^{12}$CO\,)/$L'$\,($^{13}$CO) 
and $L'$\,($^{13}$CO\,)/$L'$\,(C$^{18}$O) line luminosity ratios respectively.
Finally the Student's t-test statistical reports (t, p) to asses 
the probability that the null hypothesis 
( $L'$\,($^{12}$CO\,)/$L'$\,($^{13}$CO) and 
$L'$\,($^{13}$CO\,)/$L'$\,(C$^{18}$O) line luminosity ratios variations 
between low and high $L_{\text{IR}}$, SFR, SFE populations are not 
statistically significant) is true, located in Columns 6 and 9 
(t coefficient) and Columns 7 and 10 (probability p-value) respectively.}
\label{tbl:SplitSamplVal2}
\begin{tabular}{|l|l|l|l|l|l|l|l|l|l|l|}
\hline
\multicolumn{1}{|c|}{\multirow{2}{*}{Parameter}} &
\multicolumn{1}{|c|}{\multirow{2}{*}{Range}} &
\multicolumn{1}{|c|}{\multirow{2}{*}{N}} & 
\multicolumn{1}{|c|}{\multirow{2}{*}{Average}} &
\multicolumn{1}{|c|}{\multirow{2}{*}{$L'$($^{12}$CO)/$L'$($^{13}$CO)}} &
\multicolumn{2}{|c|}{t-test} &
\multicolumn{1}{|c|}{\multirow{2}{*}{$L'$($^{13}$CO)/$L'$(C$^{18}$O)}} &
\multicolumn{2}{|c|}{t-test} \\
& & & & & 
  \multicolumn{1}{|c|}{t} & \multicolumn{1}{|c|}{p} &
& \multicolumn{1}{|c|}{t} & \multicolumn{1}{|c|}{p} \\
\multicolumn{1}{|c|}{1} &
\multicolumn{1}{|c|}{2} &
\multicolumn{1}{|c|}{3} &
\multicolumn{1}{|c|}{4} &
\multicolumn{1}{|c|}{5} &
\multicolumn{1}{|c|}{6} &
\multicolumn{1}{|c|}{7} &
\multicolumn{1}{|c|}{8} &
\multicolumn{1}{|c|}{9} &
\multicolumn{1}{|c|}{10} \\
\hline
\hline
\\
\multirow{2}{*}{log[$L_{\text{IR}}$/L$_{\odot}$]} &
   [10.1 - 11.0]   & 13 & 10.5 $\pm$ 0.1   & 
   13.0 $\pm$ 2.5  & \multirow{2}{*}{-4.4} & \multirow{2}{*}{0.002}   &
   3.3 $\pm$ 0.8   & \multirow{2}{*}{-2.8} & \multirow{2}{*}{0.01}  \\
 & [11.1 - 11.8]   & 11 & 11.3 $\pm$ 0.1   & 
   18.8 $\pm$ 3.7  & &
  &  2.5 $\pm$ 0.6\\
\hline 
\multirow{2}{*}{SFR (M$_{\odot}${yr}$^{-1}$)} &
   [1.3  - 9.5]    & 13 &  3.7 $\pm$ 0.5   &
   13.0 $\pm$ 2.6  & \multirow{2}{*}{-4.2} & \multirow{2}{*}{0.0003}  &
    3.3 $\pm$ 0.8  & \multirow{2}{*}{-2.8} & \multirow{2}{*}{0.01}  \\
 & [10.3 - 68.5]   & 11 & 23.6 $\pm$ 4.0   & 
   18.8 $\pm$ 3.8  & &
  &  2.5 $\pm$ 0.6\\
\hline
\multirow{2}{*}{SFE (Gyr$^{-1}$)} &
    [0.4 - 0.9]    & 11 &  0.6  $\pm$ 0.1  & 
    12.4 $\pm$ 2.8 & \multirow{2}{*}{-4.8} & \multirow{2}{*}{7E-5}  &
     4.4 $\pm$ 1.2 & \multirow{2}{*}{-5.9} & \multirow{2}{*}{5E-6} \\
 & [1.1  - 11.8]   & 13 &  3.3  $\pm$ 0.6  & 
    19.4 $\pm$ 4.2 & &
 &    2.1 $\pm$ 0.5\\
\end{tabular}
\end{table*}

\begin{figure*}
\includegraphics[width=\textwidth]{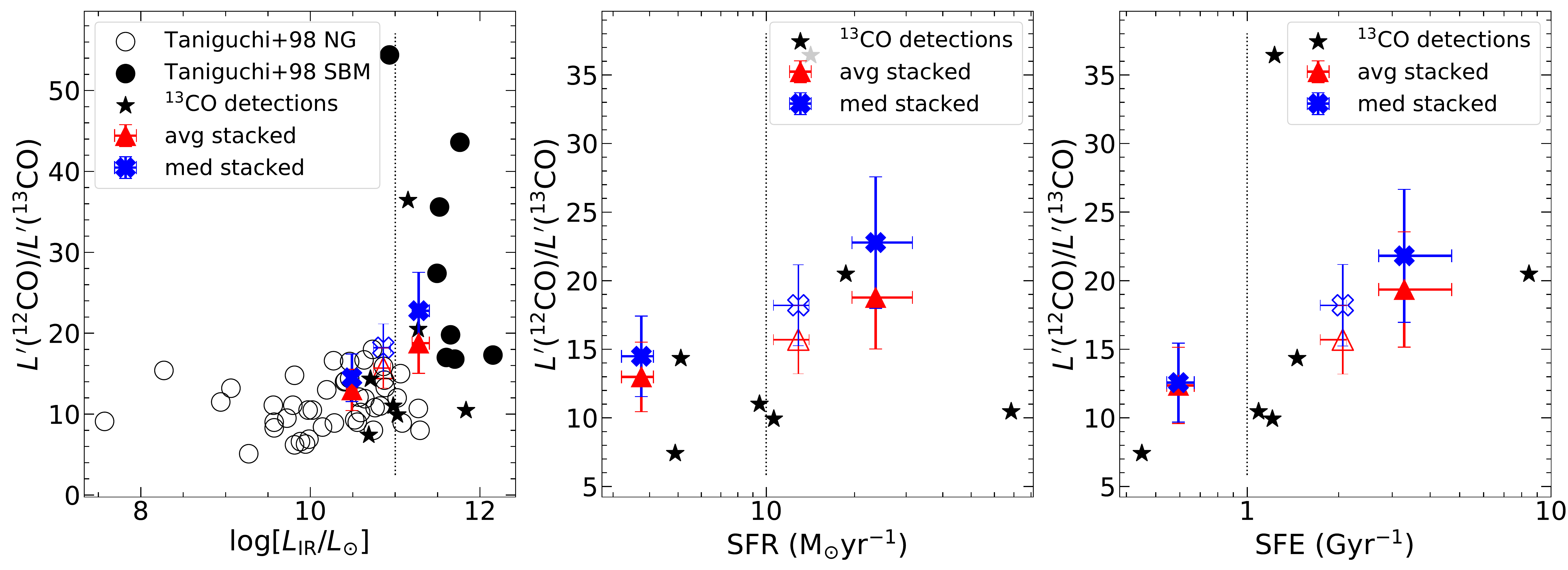}
\caption{Average (red triangles) stacked
$L'$\,($^{12}$CO\,)/$L'$\,($^{13}$CO) line luminosity ratio  for low
and high $L_{\text{IR}}$ (left panel), SFR (middle panel)  and SFE
(right panel), subsets (triangles). As reference average and median
line  luminosity ratio considering all galaxies in open symbols  and 
dashed lines indicating the boundary between low and high
$L_{\text{IR}}$,  SFR and SFE populations are included. Error bars
correspond to 1$\sigma$  confidence intervals for average (or median)
values based on Monte-Carlo simulations.
$L'$\,($^{12}$CO\,)/$L'$\,($^{13}$CO) line ratios of normal  galaxies
(open circles), starburst mergers (filled circles)  scaled by a 1.75
factor to convert from far-IR to IR luminosities  (see Appendix E from
\citealt{HerreraCamus15}).  Data from \protect\cite{Taniguchi98}  and
individual $^{13}$CO detections (stars) are also included.}
\label{fig:C18OStackResultsTrends1213} \end{figure*}

\begin{figure*}
\includegraphics[width=\textwidth]{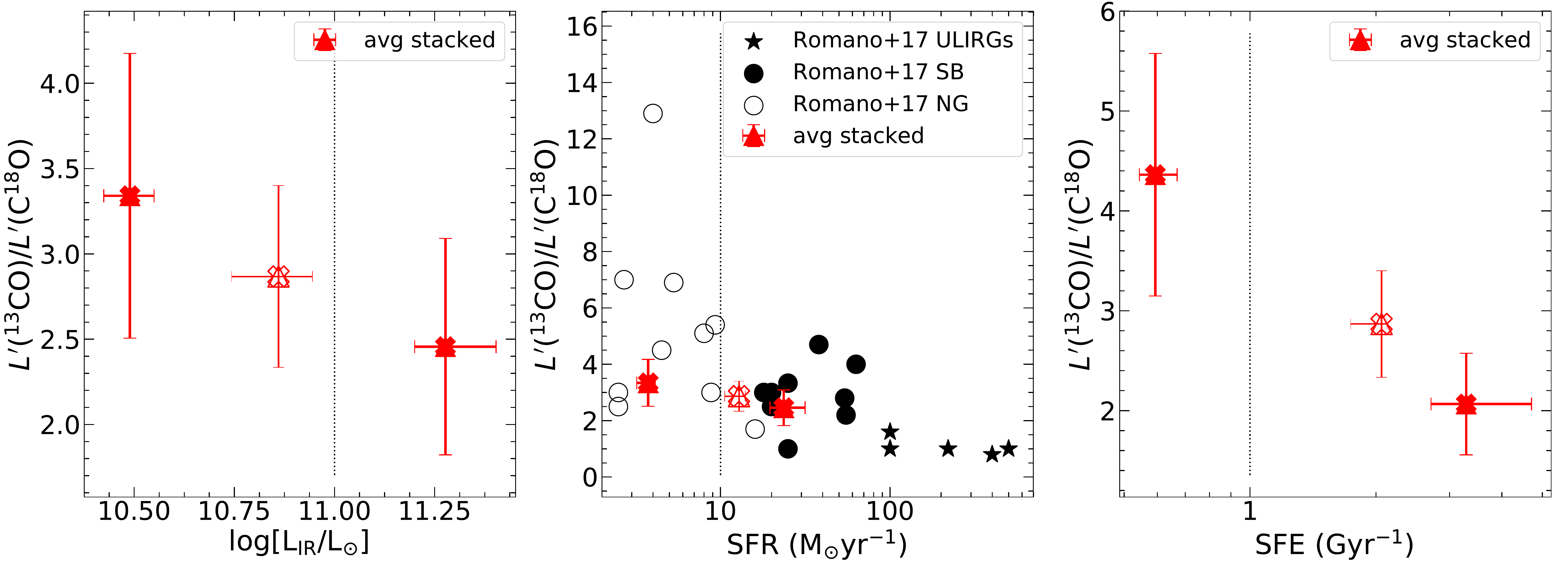}
\caption{Average (red triangles) stacked
$L'$\,($^{13}$CO\,)/$L'$\,(C$^{18}$O)  line luminosity  ratio for low
and high $L_{\text{IR}}$ (levt panel), SFR (middle panel) and  SFE
(right panel), subsets (triangles). As reference average line
luminosity  ratio considering all 24 galaxies with C$^{18}$O coverage
are shown in open  symbols and dashed lines indicating the boundary 
between low and high  $L_{\text{IR}}$, SFR and SFE populations are
included. Error bars correspond to  1$\sigma$ confidence intervals for
average values based on Monte-Carlo simulations.
$I$\,($^{13}$CO\,)/$I$\,(C$^{18}$O) ratios reported in the literature
and gathered  by \citep{Romano17} split by normal galaxies (NG open
circles), starburst galaxies  (SB filled circles) and ULIRGs (stars)
as a function of SFR are also included.  We note that median
$L'$\,($^{13}$CO\,)/$L'$\,(C$^{18}$O) line luminosities stacks are not
included in our analyses due to low significance in these values.}
\label{fig:C18OStackResultsTrends} \end{figure*}

Our ALMA Band-3 observations helped with the exploration of the
C$^{18}$O($1-0$) emission line for 24 VALES galaxies. Following a
similar approach as before, in this section we only use these 24
galaxies with simultaneous C$^{18}$O, $^{13}$CO and $^{12}$CO
observations. Figures ~\ref{fig:C18OStackResultsTrends1213} and
\ref{fig:C18OStackResultsTrends}  show the trends for 
$L'$\,($^{12}$CO\,)/$L'$\,($^{13}$CO) and
$L'$\,($^{13}$CO\,)/$L'$\,(C$^{18}$O) as a function of
$L_{\text{IR}}$, SFR and SFE respectively and  Table
\ref{tbl:SplitSamplVal2} for average
$L'$\,($^{12}$CO\,)/$L'$\,($^{13}$CO) and
$L'$\,($^{13}$CO\,)/$L'$\,(C$^{18}$O)  values. The average
$L'$\,($^{13}$CO\,)/$L'$\,(C$^{18}$O) line luminosity ratio found is
2.5$\pm$0.6, which  is in good agreement with the
$I$\,($^{13}$CO\,)/$I$\,(C$^{18}$O) line intensity  ratio found for
starburst galaxies (3.4$\pm$0.9) but slightly lower than the average
ratio found in nearby normal spiral galaxies (6.0$\pm$0.9) reported by
\cite{JimenezD17}. The central panel of Figure
\ref{fig:C18OStackResultsTrends} shows the 
$L'$\,($^{13}$CO\,)/$L'$\,(C$^{18}$O) line luminosity ratio as a
function of SFR (see Table \ref{tbl:StatSplitHighLowC18O}) and also
includes  values reported in the literature gathered by
\citealt{Romano17} and split by normal, starburts and ultra luminous
infrarred galaxies (ULIRGs).\\ 

As with $L'$\,($^{12}$CO\,)/$L'$\,($^{13}$CO) trends discussed in
Section \ref{sec:12CO13COR} we implemented a  Student's t-test to
evaluate the significance of the  
$L'$\,($^{12}$CO\,)/$L'$\,($^{13}$CO) and 
$L'$\,($^{13}$CO\,)/$L'$\,(C$^{18}$O) variations with $L_{\text{IR}}$,
SFR  and SFE considering the 24 galaxies with C$^{18}$O coverage. We
find that the differences found in the
$L'$\,($^{12}$CO\,)/$L'$\,($^{13}$CO) and 
$L'$\,($^{13}$CO\,)/$L'$\,(C$^{18}$O) variations between  low and high
$L_{\text{IR}}$, SFR, SFE populations are statistically significant
(see Table \ref{tbl:SplitSamplVal2}).  We also applied a KS test to
assess the probability that the null hypothesis (low and high: SFR,
SFE, $L_{\text{IR}}$, M$_{\text{H}_{2}}$/M$_{\sun}$,
M$_{\text{H}_{2}}$/M$_{\star}$, and  $f_{\, \text{H}_{2}}$ populations
were drawn from the same parent  population) is true, considering the
reduced 24 galaxy sample with  C$^{18}$O coverage. Table
\ref{tbl:StatSplitHighLowC18O}  shows the average values of redshift,
M$_{\text{H}_{2}}$/M$_{\sun}$, M$_{\text{H}_{2}}$/M$_{\star}$), and 
$f_{\, \text{H}_{2}}$ split by low  and high SFR values  and the
Kolmogorov-Smirnov  statistics (D, p). We find supporting evidence to
reject the null hypothesis  that the low and high SFR populations were
drawn from the same parent population. Therefore, galaxies with  high
$L'$\,($^{12}$CO\,)/$L'$\,($^{13}$CO) ratio (Figure
\ref{fig:C18OStackResultsTrends1213}) and low
$L'$\,($^{13}$CO\,)/$L'$\,(C$^{18}$O) ratio (Figure
\ref{fig:C18OStackResultsTrends})  also show relatively high
$L_{\text{IR}}$, SFR, SFE and high  reservoirs of molecular gas (see
Table~\ref{tbl:StatSplitHighLowC18O}). These line ratios can be
explained by overabundances of $^{12}$CO and C$^{18}$O (both produced
in high mass stars), with respect $^{13}$CO, that  could be understood
as a result of selective nucleosynthesis  where high-mass stars enrich
the ISM of these galaxies.

\begin{table}
\caption{Average values of different parameters (Column 1):
redshift,
molecular gas mass (M$_{\text{H}_{2}}$), molecular gas mass to 
stellar mass ratio (M$_{\text{H}_{2}}$/M$_{\star}$), and molecular 
gas fraction 
($f_{\, \text{H}_{2}}$ = M$_{\text{H}_{2}}$ / (M$_{\text{H}_{2}}$+M$_{\star}$)) 
considering 24 galaxies with C$^{18}$O coverage
split by low (Column 2) and high SFR values (Column 3) .
Columns 4 and 5 contain the Kolmogorov-Smirnov statistical 
reports (D, p) to asses the probability 
that low and high ($z$, $M_{\text{H}_{2}}$, 
$M_{\text{H}_{2}}$/$M_{\star}$, and $f_{\, \text{H}_{2}}$) populations  
were drawn from populations with 
identical distributions.} 
 \label{tbl:StatSplitHighLowC18O} 
\begin{tabular}{|l|r|r|r|r|r|r|}
\hline
  & \multicolumn{2}{c}{SFR [M$_{\odot}$yr$^{-1}$]}\\
\hline
 \multicolumn{1}{|c|}{Parameter} &
 \multicolumn{1}{c|}{low} &
 \multicolumn{1}{c|}{high} &
 \multicolumn{2}{c|}{KS}\\
 \multicolumn{1}{|c|}{} &
 \multicolumn{1}{c|}{N = 15} &
 \multicolumn{1}{c|}{N = 12} &
  \multicolumn{1}{|c|}{\multirow{2}{*}{D}} &
  \multicolumn{1}{|c|}{\multirow{2}{*}{p}} \\
 \multicolumn{1}{|c|}{} &
 \multicolumn{1}{c|}{[1.3-9.5]} &
 \multicolumn{1}{c|}{[10.3-68.5]} \\
 \multicolumn{1}{|c|}{1} &
\multicolumn{1}{|c|}{2} &
\multicolumn{1}{|c|}{3} &
\multicolumn{1}{|c|}{4} &
\multicolumn{1}{|c|}{5}\\
\hline
\hline
 <z>                                   & 0.04 $\pm$ 0.01 & 0.10 $\pm$ 0.04 & 0.8  & 7E-5 \\
 <log[M$_{\text{H}_{2}}$/M$_{\odot}$]> & 8.8  $\pm$ 0.1  &  9.9 $\pm$ 0.1  & 0.8  & 0.01 \\
 <M$_{\text{H}_{2}}$/M$_{\star}$>      & 0.15 $\pm$ 0.04 & 0.36 $\pm$ 0.05 & 0.6  & 0.02 \\
 <$f_{\, \text{H}_{2}}$>               & 0.12 $\pm$ 0.07 & 0.24 $\pm$ 0.09 & 0.6  & 0.02 \\
\end{tabular}
\end{table}

\section{Discussion}\label{sec:Disc}

The $L'$\,($^{12}$CO\,)/$L'$\,($^{13}$CO) line luminosity ratios
presented in this work can be affected by optical depth effects or by
the different physical processes that have been invoked to explain
large and low  $I$\,($^{12}$CO\,)/$I$\,($^{13}$CO) line intensity
ratios. The most relevant  ones are: i) selective photodissociation:
$^{12}$CO molecules are more abundant than  $^{13}$CO molecules and
hence due to their higher density they are self-shielded against
strong interstellar UV radiation fields,  unlike less abundant
$^{13}$CO molecules which are more easily photo-dissociated leading to
a $^{13}$CO under abundance and hence a higher
$I$\,($^{12}$CO\,)/$I$\,($^{13}$CO) line intensity ratio in regions
with  strong UV radiation fields, ii) chemical isotope-dependent
fractionation: where gas kinetic temperatures elevates $^{13}$CO
abundance through the isotopic charge exchange reaction
\citep{Watson77}, where $^{12}$CO + $^{13}$C$^{+} \rightarrow$
$^{12}$C$^{+}$ + $^{13}$CO + $\Delta$E and iii) selective
nucleosynthesis where massive stars in star  forming regions produce
significantly higher amounts of $^{12}$C   compared to $^{13}$C
leading to a high $I$\,($^{12}$CO\,)/$I$\,($^{13}$CO)  line intensity
ratio \citep{Henkel93,Aalto95}.

The higher $L'$\,($^{12}$CO\,)/$L'$\,($^{13}$CO) line luminosity ratio
found for galaxies with close companions may be explained by
interaction activity. For example, during the early stages of a merger
event, part of the  gas escapes and  disperses into the intergalactic
medium \citep{Mirabel89}. The remaining gas shrinks to the center,
becomes denser and converted  partially into molecular H$_{2}$ gas.
This fresh molecular gas with relatively low metallicity and hence a
high $^{12}$C/$^{13}$C luminosity ratio will trigger new starburst
events boosting the $^{12}$CO/$^{13}$CO abundance ratio
\citep{Casoli92b,Langer90}. On the other hand, the opposite scenario
occurs in galaxies in denser environments, like galaxy clusters, where
a deficit in $^{12}$CO is linked to the low
$I$\,($^{12}$CO\,)/$I$\,($^{13}$CO) line ratios observed. Galaxies in
clusters have lived long enough to enrich the ISM with $^{13}$C atoms
from low mass stellar nucleosynthesis, while at the same time, the
evaporation or stripping of low density GMCs as galaxies enter into
the cluster moving through the intracluster medium (ICM), possibly
reduce the presence of new starburst events and therefore would lead
to a reduced $I$\,($^{12}$CO\,)/$I$\,($^{13}$CO) intensity ratio
\citep{Alatalo15}.  In summary, the enhanced
$L'$\,($^{12}$CO\,)/$L'$\,($^{13}$CO) line luminosity ratios observed
in galaxy mergers (M) and galaxies with a projected companion (BC, DC)
which got relatively high SFR (34$\pm$6.9 and 17.6$\pm$5.9,
respectively)  could be explained by a new starburst activity in these
systems. Galaxies without any projected companion (B, D) with low SFRs
present a relatively low
$L'$\,($^{12}$CO\,)/$L'$\,($^{13}$CO)\,$=$\,3.4$\pm$0.5 line
luminosity ratio, which could be explained by $^{13}$C enrichment of
their ISM induced by low and intermediate mass stars in the  absence
of young starburst events. 

The trends of $L'$\,($^{12}$CO\,)/$L'$\,($^{13}$CO)  with
$L_{\text{IR}}$, SFR, and SFE (see Figures \ref{fig:LFIR-L-12-13CO},
\ref{fig:MLT-L-12-13CO} and \ref{fig:C18OStackResultsTrends1213})
found in this work  provide the evidence that galaxies with low SFR,
SFE and $L_{\text{IR}}$, also show low
$L'$\,($^{12}$CO\,)/$L'$\,($^{13}$CO) line luminosity ratios in
agreement to the idea that normal star forming galaxies have larger
gas consumption times to enrich with $^{13}$C the ISM from low and
intermediate mass stars. On the other hand galaxies with high SFR, SFE
and $L_{\text{IR}}$ present high $L'$\,($^{12}$CO\,)/$L'$\,($^{13}$CO)
line luminosity ratios most probably due to younger starburst
activity. These higher ratios found here are in good agreement with
scenarios in which galaxies with  higher fractions of dense molecular
gas  (see Tables \ref{tbl:StatSplitHighLow} and
\ref{tbl:StatSplitHighLowC18O}) show higher $L_{\text{IR}}$ and higher
SFE \citep{Solomon05,Solomon92}, probably  induced by the triggering
of a recent starburst episode after the in-fall of unprocessed gas to
the central galaxy regions  \citep{Henkel93,Casoli92a,Konig16}.

Similarly, the $I$\,($^{13}$CO\,)/$I$\,(C$^{18}$O) line intensity
ratios are explained  in terms of either a  $^{13}$CO deficit or an
overabundance of C$^{18}$O. Recently, low
$I$\,($^{13}$CO\,)/$I$\,(C$^{18}$O) line intensity ratios have been
reported for different type of galaxies
\cite{Danielson13,Sliwa17,JimenezD17,Brown19}.  In all these cases,
the low ratios have been attributed to the presence of  massive stars
in a recent starburst. Our trends found for 
$L'$\,($^{13}$CO\,)/$L'$\,(C$^{18}$O) (Figure
\ref{fig:C18OStackResultsTrends}) are in good agreement  to previous
ratios reported for starburst galaxies \citep{Zhang18,JimenezD17}.
Considering the 24 VALES galaxies with C$^{18}$O spectral coverage, 
we show moderate trends of $L'$\,($^{13}$CO\,)/$L'$\,(C$^{18}$O) with
$L_{\text{IR}}$, SFR, and SFE. Galaxies with higher $L_{\text{IR}}$,
SFR and SFE are found to show high
$L'$\,($^{12}$CO\,)/$L'$\,($^{13}$CO)  (Figure
\ref{fig:C18OStackResultsTrends1213}), low
$L'$\,($^{13}$CO\,)/$L'$\,(C$^{18}$O)  (Figure
\ref{fig:C18OStackResultsTrends}) luminosity ratios, and relatively
high reservoirs of molecular gas (see
Table~\ref{tbl:StatSplitHighLowC18O}). Similar results have been
associated to a top heavy IMF
(\citealt{Danielson13,Sliwa17,Zhang18,Brown19}), however in order to
break the degeneracy between young starburst and top-heavy IMF,  an
independent determination of the age of the starburst is needed.

\subsection{Optical depth, selective photodissociation and chemical
fractionation effects.}\label{}

\citet{Aalto95} pointed out the difficulties from interpreting the
$^{12}$CO and $^{13}$CO abundances from
$I$\,($^{12}$CO\,)/$I$\,($^{13}$CO)  line intensity ratios as these
might be affected by surface density,  optical depth, and gas
temperature. They suggested that the high
$I$\,($^{12}$CO\,)/$I$\,($^{13}$CO) line intensity ratios observed in
mergers and interacting galaxies \citep{Casoli92b,Henkel93} is
produced by the in-falling of unprocessed gas which could affect the
gas elemental abundances,  only if the ISM has moderate optical depths
($\tau \approx 1$). More recently, \citet{Zhang18} presented how
$^{13}$CO and C$^{18}$O  opacity affects
$I$\,($^{12}$CO\,)/$I$\,($^{13}$CO)  and
$I$\,($^{13}$CO\,)/$I$\,(C$^{18}$O)  line intensity ratios in local
thermodynamic equilibrium (LTE) and non LTE conditions assuming:
representative Galactic abundance ratios  ($^{13}$CO/C$^{18}$O =
7--10, $^{12}$CO/$^{13}$CO=70), and typical ULIRGs and SMGs 
conditions ($e.g.$ $\tau_{^{12}\text{CO}}\approx2$, $T_{kin}=30$K).
They found that the high $I$\,($^{12}$CO\,)/$I$\,($^{13}$CO) ratios
$\geq$ 30 observed in high redshift galaxies, would need extremely low
optical depths for $^{13}$CO ($\tau< 0.03$), meaning that
$I$\,($^{12}$CO\,)/$I$\,($^{13}$CO) line intensity ratios are affected
by optical depth effects. In order to properly account the optical
depths in $I$\,($^{12}$CO\,)/$I$\,($^{13}$CO) intensity ratios,
multiple line transitions observations are needed to  measure
excitation conditions and derive the optical depths of the ISM in
these galaxies, which is beyond the scope this work and hence, the
$L'$\,($^{12}$CO\,)/$L'$\,($^{13}$CO) luminosity ratios reported here 
should be taken as a  lower limit of the $^{12}$CO/$^{13}$CO abundance
ratio \citep{Henkel10,Martin19}. On the other hand, \citet{Zhang18}
found that even moderate $^{13}$CO optical depths
($\tau_{^{13}\text{CO}}$ $\sim$ 0.2-0.5) do not cause the
$I$\,($^{13}$CO\,)/$I$\,(C$^{18}$O)  line intensity ratio to deviate
significantly from more typical values ($^{13}$CO/C$^{18}$O$\sim$7),
meaning that the low $I$\,($^{13}$CO\,)/$I$\,(C$^{18}$O) found in
high-redshift starbursts and local ULIRGs reflect the intrinsic
isotopologue  abundance ratios ($i.e. $
$I$\,($^{13}$CO\,)/$I$\,(C$^{18}$O) $\approx$ $^{13}$CO/C$^{18}$O
$\approx$ $^{13}$C/$^{18}$O).

If chemical fractionation is the main physical mechanism controlling
the  observed line ratios, the  $^{13}$CO abundance would be boosted
with respect to $^{12}$CO (and C$^{18}$O) at low temperatures (T
$\approx$ 10 K \citealt{Watson76}). Nevertheless, considering a  mean
temperature T > 20 K \citep{Ibar15,Hughes17} for our VALES sample,  we
can reject chemical fractionation as the main mechanism controlling
the $L'$\,($^{12}$CO\,)/$L'$\,($^{13}$CO) ratios. On the other hand,
selective photodissociation can affect the less abundant  $^{13}$CO
and C$^{18}$O molecules compared to $^{12}$CO, however extreme
conditions with high gas densities ($>10^{26}\rm{cm}^{-3}$) are
required  \citep{Zhang18,Romano17}. However with an average gas
density of  $10^{4}\rm{cm}^{-3}$ \citep{Hughes17}, our sample of
galaxies  do not fulfill such conditions.  Moreover, knowing that
C$^{18}$O is even more sensitive to selective dissociation than
$^{13}$CO, C$^{18}$O molecules would be more dissociated than 
$^{13}$CO molecules, resulting in high
$L'$\,($^{13}$CO\,)/$L'$\,(C$^{18}$O) ratios, which is inconsistent
with results shown in Figure  \ref{fig:C18OStackResultsTrends} where
galaxies with more intense UV radiation fields associated with high
$L_{\text{IR}}$ and SFR show low 
$L'$\,($^{13}$CO\,)/$L'$\,(C$^{18}$O) ratios. Thus, the
($L'^{12}$CO)/($L'^{13}$CO) and ($L'^{13}$CO)/$L'$(C$^{18}$O) 
variations found here are not compatible with a scenario in which
selective  photodissociation or chemical fractionation play a dominant
role.

\subsection{Insights from Galactic Chemical
Evolution}\label{sec:Models}

Recently \cite{Romano17} used Galactic Chemical Evolution (GCE) models
to compute the the abundances of numerous elements including $^{12}$C,
$^{16}$O, $^{13}$C and $^{18}$O in the ISM of galaxies,  assuming that
i) stars form from raw material with primordial chemical composition,
ii) outflows remove stellar ejecta and a fraction of the surrounding
ISM, iii) star formation follows a canonical  Kennicutt-Schmidt law
\citep{Schmidt59,Kennicutt98}, iv) stars release the synthesized
elements during their lifetime, and v) stellar eject are homogeneously
mixed in the ISM, allowing to  follow multiple isotopic ratios and
trace their abundance ratios on different isotopes and different
elements. They have shown that neither selective photodissociation nor
chemical isotope-dependent fractionation can significantly perturb
globally averaged isotopologue abundance ratios, since these processes
will typically affect only small mass fractions of individual
molecular clouds in galaxies. Using these models \cite{Zhang18} was
able to compare the effects of assuming different IMF of young
starbursts by incorporating the appropriate timescales at which
different  stellar populations enrich the ISM, and conclude that a
canonical IMF  can not reproduce the observed low
$I$\,($^{13}$CO\,)/$I$\,(C$^{18}$O) ratios in  ULIRGs and SMGs. Thus
assuming that the velocity integrated line flux densities coming from
average stacks are not affected by these two other effects, we propose
that these  emission line ratios could be induced by selective
nucleosynthesis. 

\section{Conclusions}\label{Sec:Concl}

In this paper we present a stacking analysis of $^{12}$CO($1-0$),
$^{13}$CO($1-0$) emission lines of 27 galaxies,  and C$^{18}$O ($1-0$)
in 24 galaxies, belonging to the VALES survey. We have detected 6
individual $^{13}$CO (1-0) line signals from 6 galaxies in moment-0
maps, with SNR > 5. We have successfully demonstrated that it is
possible to detect the signal coming from faint emission lines
($^{13}$CO and C$^{18}$O) in low-redshift galaxies through stacking
analysis,  pushing the current ALMA detectability limits. Therefore,
the analysis presented here can be applied to detect faint signals
from different molecules coming from low, intermediate and high
redshift galaxies, exploiting radio interferometric datasets from
ALMA.

We have explored three different independent stacking analysis, two of
them in the ``image plane\,'' i) 2D-moment-0 stacking and ii) 3D-image
stacking, while a third one  in visibility's space iii) $uv$-plane 
stacking. We found that for bright emission line (as the case for
$^{12}$CO) $uv$-plane stacks produce the highest signal-to-noise
compared to 2D-moment-0, and 3D-image stacks.  Moment-zero stacked
maps for faint lines, like $^{13}$CO, shows the highest signal to
noise compared with 3D-image and $uv$-plane stackings.

We found an overall $L'$\,($^{12}$CO\,)/$L'$\,($^{13}$CO) line
luminosity ratio of 16.1$\pm$2.5. We also found a dependence of
$L'$\,($^{12}$CO\,)/$L'$\,($^{13}$CO) line luminosity ratio on optical
morphology/environment where galaxies showing a close projected
companion and mergers show boosted
$L'$\,($^{12}$CO\,)/$L'$\,($^{13}$CO) line luminosity ratios. Mergers
show at low significance a $L'$\,($^{12}$CO\,)/$L'$\,($^{13}$CO) line
luminosity ratio which is 2 times higher than that found in galaxies
without a projected companion. We also found positive trends between
$L'$\,($^{12}$CO\,)/$L'$\,($^{13}$CO) line luminosity ratio and SFR,
SFE, $L_{\text{IR}}$, $\Sigma_{\text{H}_{2}}$ and
$\Sigma_{\text{SFR}}$.

We also provide C$^{18}$O stacking analysis for 24 VALES galaxies. We
detect signal coming from the 2D moment-0 stacked images at a
significance of $\sim5\sigma$ with an average 
$L'$\,($^{13}$CO\,)/$L'$\,(C$^{18}$O) line luminosity ratio of
2.5$\pm$0.6.  This average value is in good agreement to the
$I$\,($^{13}$CO\,)/$I$\,(C$^{18}$O)\,$=$\,3.4$\pm$0.9 line ratio for
starburst galaxies found by \citet{JimenezD17}. We find negative
trends of $L'$\,($^{13}$CO\,)/$L'$\,(C$^{18}$O) as a function of
$L_{\text{IR}}$, SFR and SFE.

We recall that our L'\,($^{12}$CO\,)/L'\,($^{13}$CO) ratios can be
affected by optical depth effects, and hence  the
$L'$\,($^{12}$CO\,)/$L'$\,($^{13}$CO) luminosity ratios  reported here
should be taken as a lower limit for the $^{12}$CO/$^{13}$CO abundance
ratio. To assess  the optical depth effects on
$I$\,($^{12}$CO\,)/$I$\,($^{13}$CO) intensity  ratios, multiple line
transitions observations are needed to  measure excitation conditions
and derive the optical depths of the ISM.

Neither chemical fractionation nor selective photo-dissociation are
expected to be responsible for the trends found in this work as  the
required low temperatures ($\leq10 K$) and high densities 
($>10^{26}\rm{cm}^{-3}$) are not fulfilled by our sample of galaxies.
The combined  $L'$\,($^{12}$CO\,)/$L'$\,($^{13}$CO) and 
$L'$\,($^{13}$CO\,)/$L'$\,(C$^{18}$O) variations provide additional
evidence inconsistent with selective  photodissociation as the
responsible agents behind the results shown here.  This, leaves
selective nucleosythesis to be the most probable mechanism for the
high $L'$\,($^{12}$CO\,)/$L'$\,($^{13}$CO) and low
$L'$\,($^{13}$CO\,)/$L'$\,(C$^{18}$O) ratios found in bright far-IR
luminosity galaxies.  The scenario might be that higher molecular gas
reservoirs can trigger new starburst events where high mass  stars
enrich their ISM.

Future analyses using ALMA observations of these and other  CNO
isotoplogue molecules (C$^{17}$O, $^{12}$CN, $^{13}$CN )  on larger
samples, sampling different galaxy populations covering  wider ranges
in SFR, SFE, and molecular gas contents at different  epochs could
fill in the gap of these molecular line observations  between nearby
galaxies and lensed galaxies at high redshift, shedding  light on the
physical processes behind their star formation activity.

\section*{Acknowledgements}
We thank an anonymous referee for constructive comments and
suggestions, which helped to improve the manuscript.  HMH and EI
acknowledge partial support from FONDECYT through grant
N\textsuperscript{o}:1171710. M.J.M.~acknowledges the support of the
National Science Centre, Poland through the SONATA BIS grant
2018/30/E/ST9/00208. KKK acknowledges support from the Swedish
Research Council (2015-05580). This work also benefited from the
International Space Science Institute (ISSI/ISSI-BJ) in Bern and
Beijing, thanks to the funding of the team "Chemical abundances in the
ISM: the litmus test of stellar IMF variations in galaxies across
cosmic time" (Principal Investigator D.R. and Z-Y.Z.). TMH
acknowledges the support from the Chinese Academy of Sciences (CAS)
and the National Commission for Scientific and Technological Research
of Chile (CONICYT) through a CAS-CONICYT Joint Postdoctoral Fellowship
administered by the CAS South America Center for Astronomy (CASSACA)
in Santiago, Chile. This paper makes use of the following ALMA data:
ADS/JAO.ALMA 2013.1.00530.S. ALMA is a partnership of ESO
(representing its member states), NSF (USA) and NINS (Japan), together
with NRC (Canada), NSC and ASIAA (Taiwan), and KASI (Republic of
Korea), in cooperation with the Republic of Chile. The Joint ALMA
Observatory is operated by ESO, AUI/NRAO and NAOJ.  GAMA is a joint
European-Australasian project based around a spectroscopic campaign
using the Anglo-Australian Telescope. The GAMA input catalogue is
based on data taken from the Sloan Digital Sky Survey and the UKIRT
Infrared Deep Sky Survey. Complementary imaging of the GAMA regions is
being obtained by a number of independent survey programmes including
GALEX MIS, VST KiDS, VISTA VIKING, WISE, Herschel-ATLAS, GMRT and
ASKAP providing UV to radio coverage. GAMA is funded by the STFC (UK),
the ARC (Australia), the AAO, and the participating insti- tutions.
The GAMA website is \url{http://www.gama-survey.org/}. This research
made use of Astropy,\footnote{http://www.astropy.org}  a
community-developed core Python package for Astronomy
\citep{astropy:2013, astropy:2018}.

%%%%%%%%%%%%%%%%%%%%%%%%%%%%%%%%%%%%%%%%%%%%%%%%%%
%%%%%%%%%%% Data Availability Statement %%%%%%%%%%
\section*{Data Availability}
The data underlying this article are available in the 
ALMA Science Archive at \url{https://almascience.nrao.edu/asax/}, 
and can be accessed with the Project ID: 2013.1.00530.S.


%%%%%%%%%%%%%%%%%%%%%%%%%%%%%%%%%%%%%%%%%%%%%%%%%%
%%%%%%%%%%%%%%%%% APPENDICES %%%%%%%%%%%%%%%%%%%%%
%\appendix\label{sec:Appendix}
%If you want to present additional material which would interrupt the flow of 
%the main paper, it can be placed in an Appendix which appears after the list 
%of references.
%%%%%%%%%%%%%%%%%%%%%%%%%%%%%%%%%%%%%%%%%%%%%%%%%%

% Don't change these lines
\bsp	% typesetting comment
\label{lastpage}
\end{document}